\documentclass[prd,preprintnumbers,superscriptaddress,nofootinbib,showpacs,twocolumn]{revtex4-2}
\usepackage{pstricks}
\usepackage{color}
\usepackage{amssymb,amsmath,bm,bbm}
\usepackage{epsf,epsfig}
\usepackage{afterpage}
\usepackage{long table}
\usepackage{latexsym,mathrsfs,amssymb,dsfont}
\usepackage{graphics}
\usepackage{url}
\usepackage{paralist}
\usepackage{bbold}
\usepackage{appendix}
\usepackage{hyperref}

\newcommand{\sm}{\textsc{SM}}
\newcommand{\np}{\textsc{NP}}
\newcommand{\inter}{\textsc{INT}}
\newcommand{\be}{\begin{equation}}
\newcommand{\ee}{\end{equation}}
\newcommand{\bi}{\begin{itemize}}
\newcommand{\ei}{\end{itemize}}
\newcommand{\ba}{\begin{array}}
\newcommand{\ea}{\end{array}}
\newcommand{\bea}{\begin{eqnarray}}
\newcommand{\eea}{\end{eqnarray}}
\newcommand{\bec}{\begin{center}}
\newcommand{\eec}{\end{center}}

\newcommand{\nn}{\nonumber}

\def\@seccntformat#1{\@ifundefined{#1@cntformat}%
   {\csname the#1\endcsname\quad}
   {\csname #1@cntformat\endcsname}
}
\begin{document}

\preprint{BARI-TH/787-26}

\title{Hunting for new physics in $B$ meson $b \to u$  transitions: \\ \vspace*{0.3cm} A fresh look with Belle II data}
\author{Pietro~Colangelo}
\email[Electronic address:]{ pietro.colangelo@ba.infn.it} 
\affiliation{Istituto Nazionale di Fisica Nucleare, Sezione di Bari, via Orabona 4, 70126 Bari, Italy}
\author{Fulvia~De~Fazio}
\email[Electronic address:]{ fulvia.defazio@ba.infn.it} 
\affiliation{Istituto Nazionale di Fisica Nucleare, Sezione di Bari, via Orabona 4, 70126 Bari, Italy}
\author{Maria~Letizia~Di Cuia}
\email[Electronic address:]{ m.dicuia6@studenti.uniba.it} 
\affiliation{Dipartimento Interateneo di Fisica "Michelangelo Merlin", Universit\`a degli Studi di Bari, via Orabona 4, 70126 Bari, Italy}
\author{Davide~Milillo}
\email[Electronic address:]{ davide.milillo@ba.infn.it} 
\affiliation{Istituto Nazionale di Fisica Nucleare, Sezione di Bari, via Orabona 4, 70126 Bari, Italy}
\affiliation{Dipartimento Interateneo di Fisica "Michelangelo Merlin", Universit\`a degli Studi di Bari, via Orabona 4, 70126 Bari, Italy}

\begin{abstract}
\noindent
We present an analysis of the decays  $B^- \to \ell^- {\bar \nu}_\ell$, 
$\bar B^0 \to \pi^+ \ell^- {\bar \nu}_\ell$, and $B^- \to \rho^0 \ell^- {\bar \nu}_\ell$ leveraging  the precise measurements of binned differential branching fractions recently reported by the Belle II Collaboration for the semileptonic modes. We adopt a generalized low-energy  Hamiltonian that incorporates all dimension-six operators with left-handed neutrinos -- consistent with  the Standard Model effective field theory (SMEFT) -- featuring complex, lepton-flavor-dependent Wilson coefficients. For three representative values of $|V_{ub}|$, we perform a fit to the measured  $q^2$-binned observables to constrain the coupling parameter space and determine the corresponding confidence regions. 
Finally, we perform a  global fit for the three decay channels, determining  $|V_{ub}|$ together with the new physics couplings.

\end{abstract}

\thispagestyle{empty}


\maketitle

\section{ Introduction}
The semileptonic decays of $B$ meson play an important role  in determining fundamental parameters of the Standard Model (SM), namely the elements of  the Cabibbo–Kobayashi–Maskawa (CKM) quark mixing matrix \cite{Cabibbo:1963yz,Kobayashi:1973fv}.  Moreover,   with the advent of high-precision measurements, they provide significant tests of the SM itself, allowing  to  search for the signature  and constrain the  parameters of potential SM extensions \cite{Gambino:2020jvv}. This is the case of the modes induced by the $b \to u \ell {\bar \nu}_\ell$ transition, for  which high quality datasets are becoming available thanks to outstanding experimental efforts.

Here, we focus on the exclusive  $\bar B^0 \to \pi^+ \ell^- {\bar \nu}_\ell$ and $B^- \to \rho^0 \ell^- {\bar \nu}_\ell$ modes and on the purely leptonic  $B^- \to  \ell^- {\bar \nu}_\ell$ channel. They are  all involved in the  measurement of   the CKM element  $|V_{ub}|$, whose precise determination  remains  a long-standing challenge.  Indeed, a  tension persists  between the measurements of this parameter obtained using the
 inclusive method, studying   the decays $\bar B \to X_u \ell^- {\bar \nu}_\ell$ with no specific final state reconstructed, and the exclusive  $\bar B \to (\pi, \eta, \rho, \omega, \dots)  \ell^- {\bar \nu}_\ell$ modes  (the measurements from  $\Lambda_b$ decays are affected by larger uncertainties).
The former method yields a larger value than the latter at the level of about $2.5 \, \sigma$  \cite{HeavyFlavorAveragingGroupHFLAV:2024ctg}. This tension may be due to an  incomplete control over  theoretical uncertainties.  Alternatively, and more intriguingly, it may hint at  physics beyond the Standard Model (BSM), a possibility put forward  for the analogous  tension in  the CKM 
$|V_{cb}|$ matrix element \cite{Colangelo:2016ymy}.

Recent measurements by the Belle II Collaboration of the differential branching fractions for 
$\bar B^0 \to \pi^+ \ell^- {\bar \nu}_\ell$ and $B^- \to \rho^0 \ell^- {\bar \nu}_\ell$ decays in bins of $q^2$ (the squared momentum transfer to the lepton pair)
 provided  high-quality data to scrutinize these processes \cite{Belle-II:2024xwh}. Indeed, the availability of binned distributions enhances the sensitivity to potential non-standard effects, as the  operators in the weak effective Hamiltonian  affect the $q^2$ spectra in characteristic ways.
 \footnote{ The same applies to other semileptonic processes, such as the $D$ meson decays induced by $c \to s \mu  \nu_\mu$ \cite{Becirevic:2026tle}.}
 
To analyze the decay modes we adopt  a model-independent effective field theory approach.  We consider  the low-energy Hamiltonian describing the  $b \to u \ell {\bar \nu}_\ell$
transitions, which comprises all dimension-six operators consistent with Lorentz invariance and  left-handed neutrinos. The Hamiltonian is defined in Sec.\ref{framework}.
The Wilson coefficients weighting the various operators are treated as independent complex parameters, allowing for a broad class of  scenarios beyond  SM.
The Standard Model corresponds to the point in the parameter space where all the new Wilson coefficients vanish. In Sec.\ref{results-SM} we determine $|V_{ub}|$ at this  point using  the semileptonic distributions reported by Belle II together with  the measured branching fraction of the purely tauonic mode.
Then,  we select three representative benchmark values of $|V_{ub}|$ around such determination, and for each value  we  restrict the  new physics coefficients using  the measured  spectra
and the constraint from the purely leptonic decay $B^- \to \ell^- {\bar \nu}_\ell$.
 In this way, we can study the stability  of the  bounds under different scenarios for $|V_{ub}|$,  exploring the interplay between CKM  and  new physics (NP)   inputs. This is discussed in Secs. \ref{results-lept} - \ref{results-rho} for the three modes separately, enlightening the different  role of the semileptonic operators.  In Sec. \ref{results-combined}  we determine the best  fit of the NP parameters and their allowed regions based on  all  data for the three modes.
 Finally, we repeat the extraction of $|V_{ub}|$ in the full NP framework, allowing all NP Wilson coefficients to vary simultaneously in the fit.
The  best fit, the bounds on the NP coefficients  and their correlations indicate how to  shape the SM extensions compatible with these measurements.

\section{Framework}\label{framework}
 The analysis is based on  the  low-energy Hamiltonian 
 \begin{widetext}
 \bea
H_{\rm eff}^{b \to u \ell \nu}&=& \frac{G_F}{\sqrt{2}} V_{ub} \Big\{(1+\epsilon_V^\ell) \left({\bar u} \gamma_\mu (1-\gamma_5) b \right)\left( {\bar \ell} \gamma^\mu (1-\gamma_5) {\nu}_\ell \right) +\epsilon_R^\ell \left({\bar u} \gamma_\mu (1+\gamma_5) b \right)\left( {\bar \ell} \gamma^\mu (1-\gamma_5) {\nu}_\ell \right) \nn \\
&+&  \epsilon_S^\ell \, \left({\bar u}  b\right)  \left({\bar \ell} (1-\gamma_5) { \nu}_\ell \right) 
+ \epsilon_P^\ell \, \left({\bar u} \gamma_5 b\right)  \left({\bar \ell} (1-\gamma_5) { \nu}_\ell \right) 
+ \epsilon_T^\ell \, \left({\bar u} \sigma_{\mu \nu} (1-\gamma_5) b\right) \,\left( {\bar \ell} \sigma^{\mu \nu} (1-\gamma_5) { \nu}_\ell \right) \Big\} \nn \\ 
&+& h.c. \label{heff} 
\eea
\end{widetext}
comprising  the full set of dimension 6 semileptonic operators with left-handed neutrinos. This  is the general expression deriving in the Standard Model effective field theory (SMEFT) framework \cite{Grzadkowski:2010es}. $G_F$  is the Fermi constant. The  complex Wilson coefficients  $\epsilon^\ell_{V,R,S,P,T}$  are  lepton-flavour dependent. 
The SM corresponds to the point   $\epsilon^\ell_{V,R,S,P,T}=0$ in the space of the  coefficients.  
The observables relative to the  decay modes  governed by the Hamiltonian  \eqref{heff}  exhibit a different  sensitivity to the various  operators. This is  a crucial  feature for pinning down the coefficients.

\subsection{$ B^- \to \ell^- \bar \nu_\ell$}
We start from the $B^-$ purely leptonic decay mode.
The decay  width resulting from \eqref{heff} reads \footnote{Electromagnetic and strong radiative corrections for this mode are discussed  in \cite{Cornella:2026zkd}.}
\bea
&&\Gamma(B^- \to  \ell^- \bar \nu_\ell)=\frac{G_F^2 |V_{ub}|^2f_B^2m_B^3}{8\pi }\left(1-\frac{m_\ell^2}{m_B^2} \right)^2 \quad \nn \\
&&\qquad \,  \times \left|  \left(\frac{m_\ell}{m_B} \right)  (1+\epsilon_V^\ell- \epsilon_R^\ell)+\frac{m_B}{m_b+m_u}\epsilon_P^\ell \right|^2  . \label{Blnu}
\eea
The  decay constant $f_B$  parametrizes the current-vacuum matrix element
\be
\langle 0|{\bar u} \gamma_\mu \gamma_5 b |  B^- (p)\rangle= i \, f_B p_\mu \,\,\, ,
\ee
with $p$ the $B^-$  momentum.
The axial Ward identity has been used to relate  the matrix element of the pseudoscalar and axial-vector quark currents.
The   quark current masses are set to $m_b=4.183$ GeV and $m_u=2.2$ MeV. 
Eq.~\eqref{Blnu} shows that this mode gives access to the   Wilson coefficients
\be
 \epsilon_2^\ell=\epsilon_V^\ell-\epsilon_R^\ell \,\,\,, \qquad \epsilon_P^\ell \,\,\,  ,  \, \label{epsLept}
\ee
and that the pseudoscalar operator lifts the chiral suppression. 
The only hadronic input is the decay constant $f_B$, for which  we use the lattice QCD result 
$f_B=190\pm 1.3$ MeV   \cite{FlavourLatticeAveragingGroupFLAG:2024oxs}.
\\ 

\subsection{ $\bar{B}^0 \to \pi^+ \ell^- \bar{\nu}_{\ell}$}
The  $\bar{B}^0 \to \pi^+ \ell^- \bar{\nu}_{\ell}$ decay distribution in $q^2$,  the  squared dilepton invariant mass, resulting from the effective Hamiltonian   \eqref{heff} is given by \cite{Colangelo:2019axi}
\begin{widetext}
\bea
&&
\frac{d\Gamma}{dq^2}(\bar{B}^0 \to  \pi^+ \ell^- \bar{\nu}_{\ell}) =\frac{G_F^2 |V_{ub}|^2 \lambda^{1/2}}{128 m_B^3 \pi^3 q^2 } \left( 1 - \frac{m_\ell^2}{q^2} \right)^2 
\Bigg\{ \left| m_\ell (1 + \epsilon_V^\ell+ \epsilon_R^\ell)  +  \frac{q^2 \epsilon_S^\ell}{m_b-m_u} \right|^2 (m_B^2 - m_\pi^2)^2 f_0^2(q^2) \nn \\
&+&  \lambda  \Bigg[ \frac{1}{3} \left| m_\ell (1 + \epsilon_V^\ell+ \epsilon_R^\ell) f_+(q^2) + \epsilon_T^\ell \frac{4 q^2 f_T(q^2)}{m_B+m_\pi} \right|^2
+ \frac{2 q^2}{3} \left| (1 + \epsilon_V^\ell+ \epsilon_R^\ell) f_+(q^2) +4 \epsilon_T^\ell  \frac{m_\ell f_T(q^2)}{m_B+m_\pi} \right|^2 \Bigg] \Bigg\} \,\, . \label{Bpilnu} 
\eea
\end{widetext}
$\lambda\equiv  \lambda(m_B^2,m_\pi^2,q^2) $ is  the K\"all\'en function. The kinematical range of $q^2$ is $ [m_\ell^2, (m_B-m_\pi)^2]$. Eq.~\eqref{Bpilnu} is obtained  using the parametrization of the $\bar B^0 \to \pi^+$ matrix elements of the quark currents in terms of form factors  $f_i \equiv f_i^{B \to \pi}$ given in   Appendix~\ref{app-ff}.  The  relation between $f_S^{B \to \pi}$ and $f_0^{B \to \pi}$ given in the Appendix is also exploited.  Eq.~\eqref{Bpilnu} shows that  this mode gives access to the  coefficients
\be
\epsilon_1^\ell=\epsilon_V^\ell+\epsilon_R^\ell \,\,\, , \hskip 0.8cm \epsilon_S^\ell  \,\,\, , \hskip 0.8 cm \epsilon_T^\ell \,\,\, .
\label{epsPi}
\ee
The pseudoscalar operator  in \eqref{heff}  does not contribute due to parity conservation. The scalar and tensor operators affect the large $q^2$ range of the distribution.

In our analysis we adopt the form factors obtained in  Ref.~\cite{Leljak:2021vte},  where the Bourrely-Caprini-Lellouch (BCL) parametrization \cite{Bourrely:2008za} was implemented together with a suitable modification of the scalar form factor. The  covariance matrix for the form factor parameters is provided in the same reference.
\subsection{$ B^- \to \rho^0 \ell ^-  {\bar \nu}_\ell$}
When $\rho^0$ decays to two pions,   the expression of the fully differential decay width obtained from  the generalized low-energy Hamiltonian \eqref{heff} 
$\displaystyle \frac{d \Gamma (B^- \to \rho^0(\to \pi^+ \pi^-) \ell^-  {\bar \nu}_\ell}{ d q^2 d \cos \theta_V \, d \cos \theta \, d \phi}$
involves  a set of ten angular coefficient functions  $I_i(q^2)$. The definition of the angles $\theta_V$, $\theta$ and $\phi$ and the expression of the fully differential decay distribution  
 can be found in \cite{Colangelo:2019axi,Colangelo:2024mxe}.  We use the parametrization given  in 
 Appendix \ref{app-ff}  for the $B \to \rho$ matrix elements of the various quark currents  in terms of form factors.   The angular coefficient functions
 depend on  the  form factors  and on  the  Wilson coefficients $\epsilon_i^\ell$. After integrating over the angular variables,   four angular coefficient functions  $I_{1s},\,I_{1c},\,I_{2s}$ and $I_{2c}$  determine the $q^2$ distribution, which reads 
 \be
 \frac{d\Gamma}{dq^2}(B^- \to \rho^0 \ell ^-  {\bar \nu}_\ell) ={\cal C}_\rho |{\vec p}_\rho| \Big[3I_{1c}+6I_{1s}-I_{2c}-2I_{2s}\Big] \,\,\, . \label{dGdq2rho}
 \ee
$ {\cal C}_\rho$ is defined as
 \be
  {\cal C}_\rho = \displaystyle\frac{G_F^2|V_{ub}|^2}{1536 \pi^3 m_B^2}\left(1-\frac{m_\ell^2}{q^2} \right)^2   \label{Crho}
 \ee
 and
 \be
 |{\vec p}_\rho|=\frac{\lambda^{1/2}(m_B^2,m_\rho^2,q^2)}{2m_B}  \label{vecp}
 \ee
is  the $\rho^0$ three-momentum in the $B$ rest frame. 
 The expressions of  $I_{1c}, I_{1s}, I_{2c}$ and $ I_{2s}$  are given  in Appendix~\ref{appcoeff}. 
 The isospin factor $1/2$ has been  included in \eqref{Crho} to account for the final $ \rho^0$.

Considering the coefficient functions one finds   that the $q^2$ distribution  gives access to 
\be
\epsilon_1^\ell=\epsilon_V^\ell+\epsilon_R^\ell\,\, , \,\,\,\,  \epsilon_2^\ell=\epsilon_V^\ell-\epsilon_R^\ell\,\, , \,\,\,\, \epsilon_P^\ell\,\, , \,\,\,\, \epsilon_T^\ell \,\,\,, \label{epsRho}
\ee
since the scalar operator does not contribute.
In our analysis we use the $ \bar B^0 \to \rho^+$ form factors  computed in \cite{Bharucha:2015bzk} using Light-Cone QCD sum rules constrained by lattice QCD data, and we exploit isospin symmetry.  The full error correlation matrix for the parameters of all form factors is  provided in the same reference.
\\

Eqs.~\eqref{epsLept}, \eqref{epsPi} and \eqref{epsRho} show the different sensitivity  of the three decay modes to the operators in the generalized low-energy Hamiltonian. This is the basis of our analysis. The complementarity of the three channels plays  an essential role for isolating potential NP effects and constraining the new Wilson coefficients.

\section{Data, analyses,  results}\label{results}

For the $\bar B^0 \to \pi^+$ and $B^- \to \rho^0$ semileptonic modes we use the  Belle II Collaboration data obtained for the integrated luminosity  $364\,\mathrm{fb}^{-1}$ \cite{Belle-II:2024xwh}. Although  measurements of the decay distributions are available also from the same and other experiments \cite{Belle:2010hep,BaBar:2010efp,BaBar:2012thb,Belle:2013hlo} and \cite{CLEO:1999mif,CLEO:2007vpk,Belle:2006hlt} (with results collected in the corresponding sections in \cite{HeavyFlavorAveragingGroupHFLAV:2024ctg}), we restrict our analysis to this dataset only. This guarantees that both channels are measured within a single, homogeneous experimental setup, ensuring a consistent treatement of the systematic uncertainties across the two modes. 

The  differential $\bar B^0 \to \pi^+ \mu^- \bar\nu_\ell$ and $B^- \to \rho^0 \mu^- \bar\nu_\ell$  branching fractions\footnote{The  measurements actually refer to the average of the muon and electron sample. At the level relevant for our analysis, we do not distinguish between electrons and muons, and use the experimental data for the analysis of the muon case.}  are measured in bins of $q^2$

\begin{equation}
\Delta \mathcal{B}_i = \int_{q^2_i}^{q^2_{i+1}} dq^2 \, \frac{d\mathcal{B}}{dq^2} \, 
\end{equation}
covering the full kinematic range.
Specifically, for  $\bar B^0 \to \pi^+ \ell^- \bar\nu_\ell$ the measurements are performed in $13$ bins $[q^2_i,\,q^2_{i+1}]$,  while for $B^- \to \rho^0 \ell^- \bar\nu_\ell$ the measurements are carried out in  $10$ bins of the respective kinematic ranges. The statistical and systematic uncertainties are provided for each bin of $q^2$, together with the full error correlation matrices  \cite{Belle-II:2024xwh}.

The experimental covariance matrix $C_{\rm exp}$ is constructed combining statistical and systematic contributions  and  accounting for bin-to-bin correlations.
We include the  theoretical uncertainties associated with the hadronic inputs in the decay amplitudes,  the form factors. The uncertainties are propagated to the binned observables and encoded in a  covariance matrix $C_{\rm th}$  accounting for  the uncertainties and the correlations among different $q^2$ bins. The full covariance matrix is the sum
\begin{equation}
C = C_{\rm exp} + C_{\rm th} \, . \label{covtot}
\end{equation}
The binned theoretical  branching fractions are obtained  integrating the distributions  in Sec.~II over  each $q^2$ bin:
\begin{equation}
\Delta \mathcal{B}_i^{\rm th}(\epsilon^\mu_j, |V_{ub}|) = 
\int_{q^2_i}^{q_{i+1}^2} dq^2 \, \frac{d\mathcal{B}}{dq^2}(q^2, \epsilon^\mu_j, |V_{ub}|)\, ,
\end{equation}
where $\epsilon^\mu_j$ is the set of NP couplings for the muon.

To constrain  the parameter space of the  coefficients $\epsilon_i^\mu$ we use two  methods.

The first procedure starts from the $\chi^2$ function
\begin{equation}
\chi^2= 
\left( \Delta \mathcal{B}^{\rm exp} - \Delta \mathcal{B}^{\rm th} \right)^T
C^{-1}
\left( \Delta \mathcal{B}^{\rm exp} - \Delta \mathcal{B}^{\rm th} \right)\, .
\end{equation}
$C$  is  the covariance matrix  in \eqref{covtot} and $ \Delta \mathcal{B}^{\rm exp} - \Delta \mathcal{B}^{\rm th}$   a vector with components  $\Delta \mathcal{B}^{\rm exp}_i - \Delta \mathcal{B}^{\rm th}_i$, the index $i$ running over the number of bins in each channel.
We then construct the likelihood function $\cal L$  
\be
{\ln}[{\cal L}]=-\frac{1}{2} \big(\chi^2+ {\ln }[{\rm det}(C)] \big) + const \,\,. \label{like}
\ee
 The covariance matrix depends on the $\epsilon_j$ parameters and cannot be included  in the constant.
We define the function 
\be
\chi^2_{\rm eff}=\chi^2+{\ln } [{\rm det}(C)]
\ee
and determine $\chi^2_{\rm eff,min}$.
Computing  
\begin{equation}
\Delta \chi^2 = \chi^2_{\rm eff} - \chi^2_{\rm eff, min} \, 
\end{equation}
we determine  confidence regions for $n$ parameters by the condition 
\begin{equation}
\Delta \chi^2 \leq \Delta \chi^2_{\rm CL}(n)\, .\label{constraint}
\end{equation}
$\Delta \chi^2_{\rm CL}(n)$ corresponds to the selected  confidence level.
The real and imaginary parts of  the couplings   are treated as independent parameters.
To speed up the search for the minimum and  of the allowed regions, 
we implemented a supervised machine-learning algorithm, an artificial intelligence tool available within the  Wolfram \textit{ Mathematica} framework based on a Random Forest regression model. We start by defining the allowed ranges of variation for all NP parameters, which are independently sampled within  the interval $[-1,1]$.
 The parameter space is first constrained  by requiring  the experimental branching fractions to be reproduced within $2\sigma$. 
Within this region, we  identify an initial  set of parameter points  satisfying the condition \eqref{constraint}. 
This dataset is used to train a Wolfram \textit{ClassifierFunction} based on the Random Forest model. A large set of randomly generated parameter points is then evaluated by the classifier and assigned to either the admissible or non-admissible class according to condition \eqref{constraint}  choosing the appropriate  $\Delta \chi^2$ corresponding to $1 \sigma$ C.L. The points classified as admissible are subsequently subjected to an explicit verification procedure in order to remove possible false positives.

\begin{figure*}[t]
\begin{center}
\includegraphics[width = 0.27\textwidth]{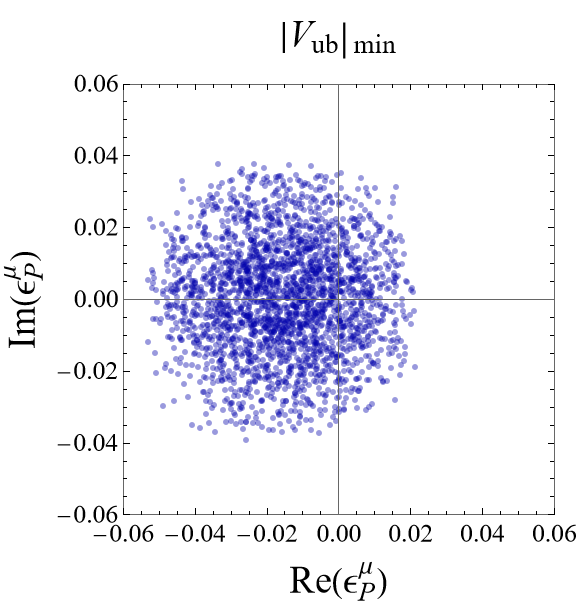} \hspace*{0.2cm}
\includegraphics[width = 0.27\textwidth]{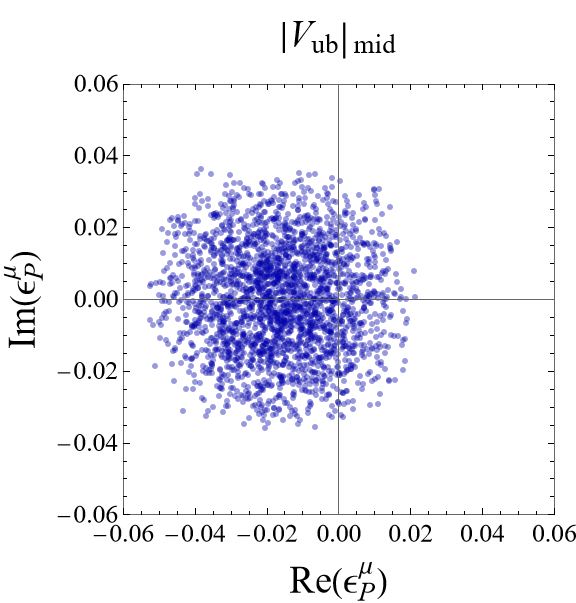} \hspace*{0.2cm}
\includegraphics[width = 0.27\textwidth]{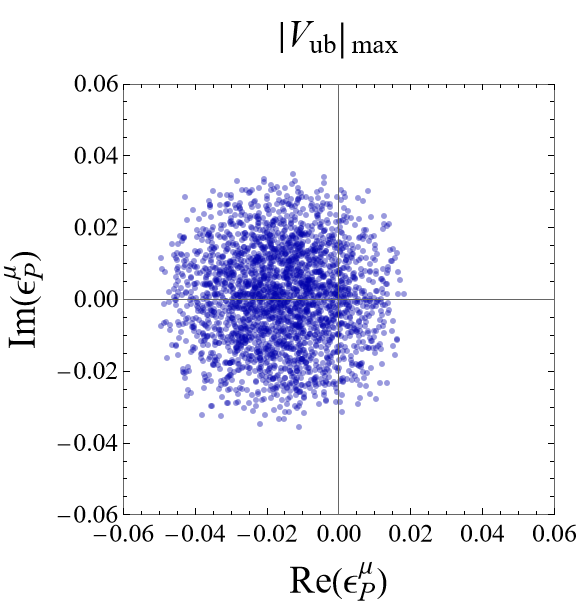}
\caption{\baselineskip 10pt  \small  Real and imaginary parts of the Wilson coefficient $\epsilon_P^\mu$ constrained by the upper bound \eqref{upper} for  the $B^- \to \mu^- \bar \nu_\mu$ decay rate.  The three benchmark values of $|V_{ub}|$ are separately considered.}\label{figlept}
\end{center}
\end{figure*}

The second procedure  to select the best values and  the  allowed regions in the Wilson coefficient  parameter  space is based on a  Monte Carlo method \cite{Trotta_2008,ParticleDataGroup:2024cfk}. Considering as an example  $\bar{B}^0\to \pi^+ \ell^- \bar{\nu}_\ell$, we assume that the probability distribution describing the experimental data in \cite{Belle-II:2024xwh} depends on the set $\boldsymbol{\theta}=\{\text{Re}(\epsilon_1^{\ell}),\text{Im}(\epsilon_1^{\ell}),\text{Re}(\epsilon_S^{\ell}),\text{Im}(\epsilon_S^{\ell}),\text{Re}(\epsilon_T^{\ell}),\text{Im}(\epsilon_T^{\ell})\}$ of   NP parameters  treated as random variables. The available information about $\boldsymbol{\theta}$ is encoded in the joint posterior probability density function (p.d.f.) 
\begin{equation}
    p(\boldsymbol{\theta}|\text{data})\propto \mathcal{L}(\boldsymbol{\theta})\pi(\boldsymbol{\theta}) \, ,
\label{bayes}
\end{equation}
where $\mathcal{L}(\boldsymbol{\theta})=p(\text{data}|\boldsymbol{\theta})$ denotes the likelihood defined in \eqref{like} and $\pi(\boldsymbol{\theta})$ the prior density functions (the   normalization constant is irrelevant). The results of inference on a selected  $(\text{Re}(\epsilon_{j}^{\ell}), \text{Im}(\epsilon_{j}^{\ell}))$  pair are summarized by the marginal posterior density function,
\bea
    p(\text{Re}(\epsilon_{j}^{\ell}), \text{Im}(\epsilon_{j}^{\ell})|\text{data}) \hspace*{3cm} \nn \\
    =\int p\bigl(\text{Re}(\epsilon_{j}^{\ell}), \text{Im}(\epsilon_{j}^{\ell}), \boldsymbol{\phi}|\text{data}\bigr) d \boldsymbol{\phi} \, , \quad 
\label{marginal}
\eea
with  $\boldsymbol{\phi}$ the set of remaining parameters. Since for the obtained multivariate distribution marginalization is not analytically accessible,  we employ the Markov Chain Monte Carlo (MCMC) method based on the Metropolis-Hastings algorithm, which generates random points $\boldsymbol{\theta}$ distributed as 
$p(\boldsymbol{\theta}|\text{data})$.

Without further constraints, we conservatively assume that the parameters $\boldsymbol{\theta}$  can take  values within the range $(-1,1)$,  each other independent:
\begin{equation}
    \pi (\boldsymbol{\theta})= \prod_{j=1}^6 \pi(\theta_j), \quad \pi(\theta_j)=\begin{cases} 
      1 & |\theta_j|<1, \\
      0 & \text{otherwise,}
   \end{cases}
\label{prior}
\end{equation}
 $\theta_j$ being any one of the NP parameters.
To map out the joint $p(\boldsymbol{\theta}|\text{data})$ p.d.f. posterior, we first  choose an arbitrary starting point $\boldsymbol{\theta}_0$ for the Markov chain, 
and a symmetric proposal density $q(\boldsymbol{\theta}|\boldsymbol{\theta}_0)$ from which independent random values of $\boldsymbol{\theta}$ are sampled. We set $\boldsymbol{\theta}_0 = (0,...,0)$ and choose for $q(\boldsymbol{\theta}|\boldsymbol{\theta}_0)$ a uniform distribution of width $\delta$,  centered at $\boldsymbol{\theta}_0$. The Metropolis generation is then carried out, and is repeated after changing the starting point.
The proposal width $\delta$ is tuned to adjust the total acceptance rate of the Monte Carlo sampling
to the optimal value of about $30\%$ \cite{10.1214/ss/1015346320}.\footnote{While a uniform width $\delta=0.3$ across all directions of parameter space is suitable in case of  $\bar{B}^0\to \pi^+ \ell^- \bar{\nu}_\ell$, for  $B^-\to \rho^0 \ell^- \bar{\nu}_\ell$ the  constraints on the parameter space require shrinking the width along the $\bigl(\text{Re}(\epsilon_{P}^{\ell}),\text{Im}(\epsilon_{P}^{\ell})\bigr)$ directions to $\delta_P=0.01$, while leaving the others unchanged.} 
The best-fit point $\hat{\boldsymbol{\theta}}=\arg\max_{\boldsymbol{\theta}} \mathcal{L}(\boldsymbol{\theta})$ is found as the one corresponding to the maximum value of $\mathcal{L}$ among those sampled by the chains.
Marginalization over the subset $\bigl(\text{Re}(\epsilon_{j}^{\ell}),\text{Im}(\epsilon_{j}^{\ell})\bigr)$ is performed by simply ignoring  the remaining coordinates. 

The numerically mapped posterior density  can be used to determine approximate highest posterior density (HPD) regions. A $(1-\alpha)\times100\%$ HPD region is the smallest region of parameter space having the $(1-\alpha)$ probability of containing the true value of the  parameters. In the $\epsilon_{j}^{\ell}$ complex plane, it is defined as 
\begin{equation}
    \mathcal{R}_{\alpha}^{(j)}=\Bigl\{
    \bigl(\text{Re}(\epsilon_{j}^{\ell}),\text{Im}(\epsilon_{j}^{\ell})\bigr) \ \bigl| \ p\bigl(\text{Re}(\epsilon_{j}^{\ell}),\text{Im}(\epsilon_{j}^{\ell})\big|\text{data}\bigr)\geq p_{\alpha}
    \Bigr\},
\label{hpd}
\end{equation}
where the threshold $p_{\alpha}$ is the largest number such that
\begin{equation}
    \iint_{\mathcal{R}_{\alpha}^{(j)}}
    p\bigl(\text{Re}(\epsilon_{j}^{\ell}),\text{Im}(\epsilon_{j}^{\ell})\, \big|\text{data}\bigr) \, d \text{Re}(\epsilon_{j}^{\ell}) \, d \text{Im}(\epsilon_{j}^{\ell})=1-\alpha.
\label{alphaint}
\end{equation}
The resulting  $p\bigl(\text{Re}(\epsilon_{j}^{\ell}),\text{Im}(\epsilon_{j}^{\ell})\big|\text{data}\bigr)$ surface is approximated by a continuous function 
$k\bigl(\text{Re}(\epsilon_{j}^{\ell}),\text{Im}(\epsilon_{j}^{\ell})\bigr)$ reconstructed from the whole set of sampled points through kernel density estimation (KDE).\footnote{This is  implement   using the  Wolfram \textit{Mathematica} function \texttt{SmoothKernelDistribution}  \cite{Wolfram:Doc}.} The threshold value $p_{\alpha}$ is found numerically as the $(1-\alpha)^{\text{th}}$ quantile of the function values $k\bigl(\text{Re}(\epsilon_{j}^{\ell}),\text{Im}(\epsilon_{j}^{\ell})\bigr)$ computed over the set of sampled points. This ensures that the probability  integrated over the region bounded by the contour $k\bigl(\text{Re}(\epsilon_{j}^{\ell}),\text{Im}(\epsilon_{j}^{\ell})\bigr)=p_{\alpha}$ is  $(1-\alpha)$, as required by Eq.~\eqref{alphaint}. 

 The two methods identify consistent best-fit solutions and similar allowed regions in the space of the Wilson coefficients. Quantitatively, the confidence regions obtained from the profile-likelihood analysis based on the $\Delta\chi^2$ criterion are  slightly broader than the corresponding HPD regions derived from the MCMC analysis. This difference is expected, since the two approaches rely on different statistical constructions. The profile-likelihood method retains all parameter configurations satisfying the condition \eqref{constraint}, whereas the second approach is based on marginal posterior distributions obtained after integrating over the remaining NP parameters. The marginalization suppresses extended low-probability directions in the multidimensional parameter space, yielding more localized HPD regions. Consequently, the $68\%$ profile-likelihood regions are often comparable in extent to  HPD regions at higher credibility levels. In the following, we present the results obtained with the MCMC approach, and the profile-likelihood results are used as an independent cross-check.

\subsection{Standard Model and $|V_{ub}|$}\label{results-SM}
Assuming the Standard Model, with all NP coefficients in \eqref{heff} set to zero, we  determine  $|V_{ub}|$ from the Belle II semileptonic data and the measured branching fraction of the purely tauonic $B^-$ decay mode:
${\cal B}(B^- \to \tau^- \bar \nu_\tau)= (1.09 \pm 0.24) \times 10^{-4}$ \cite{ParticleDataGroup:2024cfk}. The result is
 \be
 |V_{ub}|=\left(3.84^{+0.18}_{-0.17} \right) \times 10^{-3} . \label{VubSM}
 \ee 
 The pionic mode provides the tightest individual constraint on $|V_{ub}|$.
 
 We use this result  as the central benchmark value $|V_{ub}|_{\rm mid}$ within a triad of values where the remaining two are  $|V_{ub}|_{\rm min}= 3.51 \times 10^{-3}$ and $|V_{ub}|_{\rm max}=4.19 \times 10^{-3}$. 
   The last value  is in the highest range of the results obtained from the $B$  inclusive  semileptonic decay rate
    \cite{HeavyFlavorAveragingGroupHFLAV:2024ctg}.  We evaluate  the allowed regions of the new Wilson coefficients separately  for the  three benchmark values: This  allows us   to shed light on the interplay between the CKM matrix element and the couplings  $\epsilon^\mu_j$.

\subsection{$B^- \to \mu^- \bar{\nu}_{\mu}$}\label{results-lept}
For the  purely muonic decay, the upper limit  \cite{ParticleDataGroup:2024cfk}
\be
{\cal B}(B^- \to \mu^- \bar \nu_\mu)<8.6 \times 10^{-7} \,\,\, ({\rm 90 \%\, C.L.}) \label{upper}
\ee
tightly bounds the coefficient $\epsilon_P^\mu$, leaving the difference $\epsilon_2^\mu=\epsilon_V^\mu-\epsilon_R^\mu$   less constrained.
 The  allowed regions of $\epsilon_P^\mu$  compatible with \eqref{upper}, obtained by Monte Carlo rejection sampling,   are depicted in Fig.~\ref{figlept}  for the three benchmark values of $|V_{ub}|$.   Allowing   both  the real and imaginary parts of   $\epsilon_2^\mu$ to vary  in the conservative range $[-1,1]$ the bound \eqref{upper} is still satisfied. The regions in the complex $\epsilon_P^\mu$ plane are asymmetric with respect to the ${\rm Re}(\epsilon_P^\mu)=0$ axis, being shifted towards negative values of  ${\rm Re}(\epsilon_P^\mu)$. Moreover, they are almost insensitive to the value of $|V_{ub}|$. Both these features  can be  understood considering Eq.~\eqref{Blnu}. 
 
 It is worth examining if a coefficient $\epsilon_P^\tau$ of this size is compatible with the measured purely tauonic $B^-$ decay rate. The answer is in the affirmative, although it requires  some fine-tuning. Indeed, in that case a cancellation can occur in Eq.~\eqref{Blnu} between the term proportional to $\epsilon_2^\tau$  and the term controlled by 
 $\epsilon_P^\tau$.

\subsection{$\bar{B}^0 \to \pi^+ \mu^- \bar{\nu}_{\mu}$}\label{results-pi}

\begin{figure*}[t!]
\begin{center}
\includegraphics[width = 0.23\textwidth]{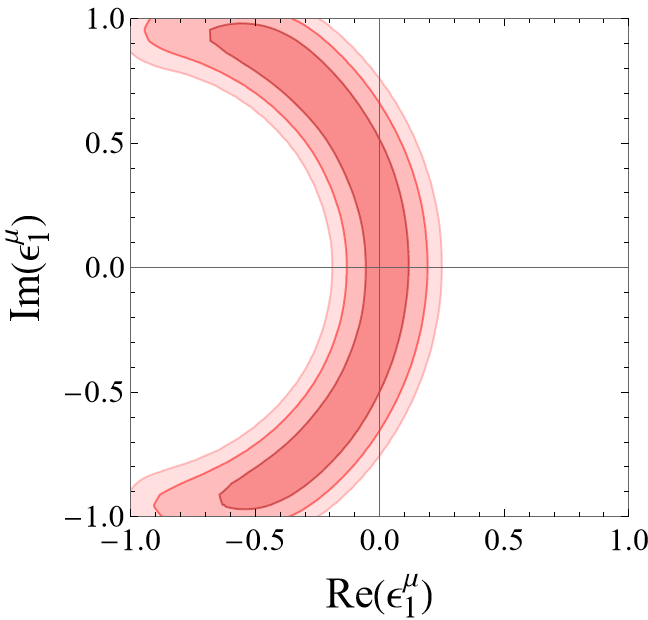} 	  \hspace*{0.2cm}
\includegraphics[width = 0.23\textwidth]{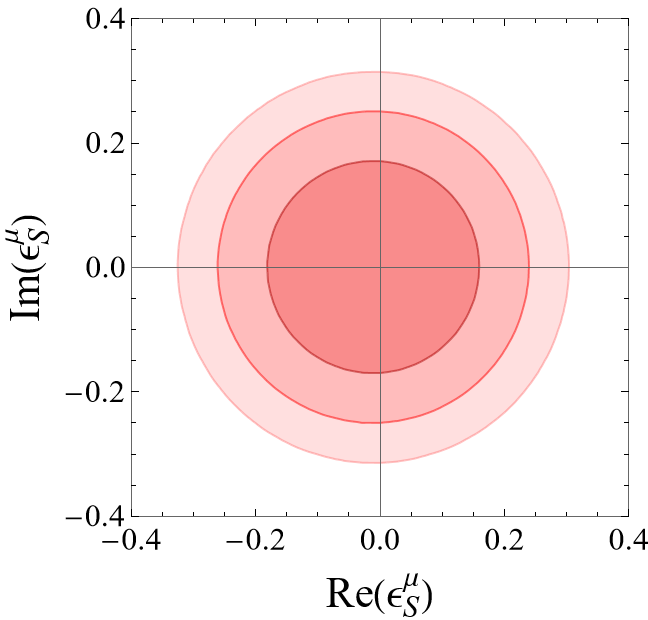}     \hspace*{0.2cm}
\includegraphics[width = 0.23\textwidth]{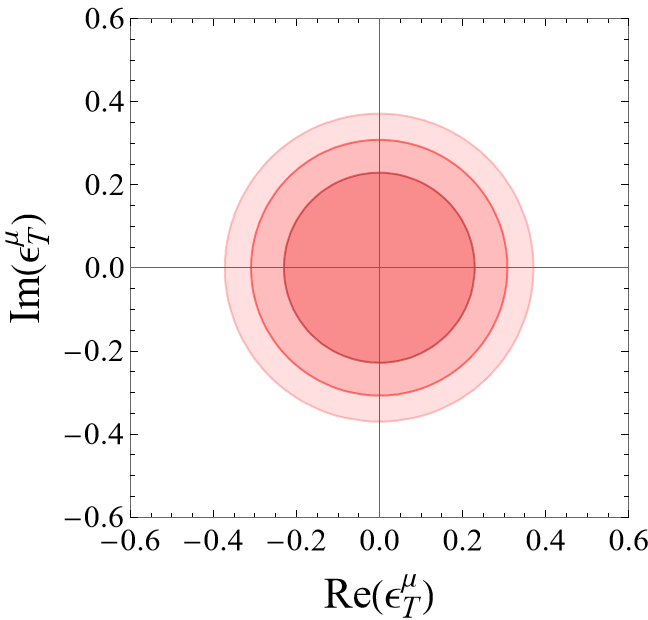}  \\  	\vspace*{0.1cm}
\includegraphics[width = 0.23\textwidth]{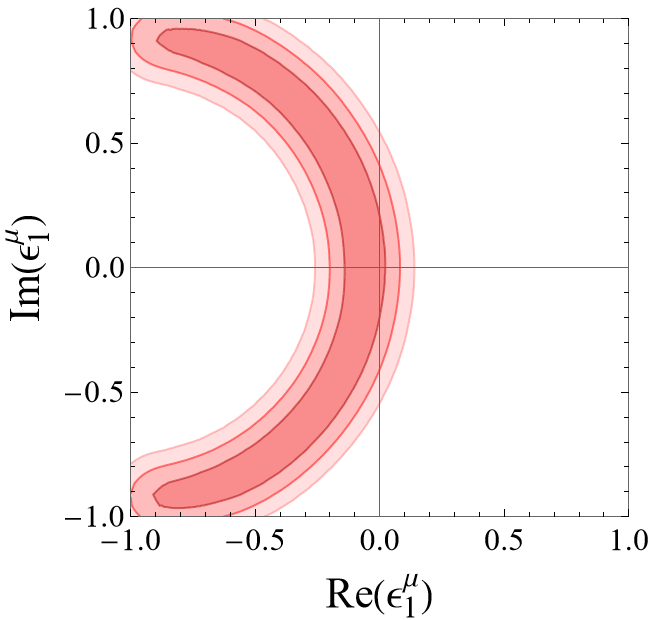} 	   \hspace*{0.2cm}
\includegraphics[width = 0.23\textwidth]{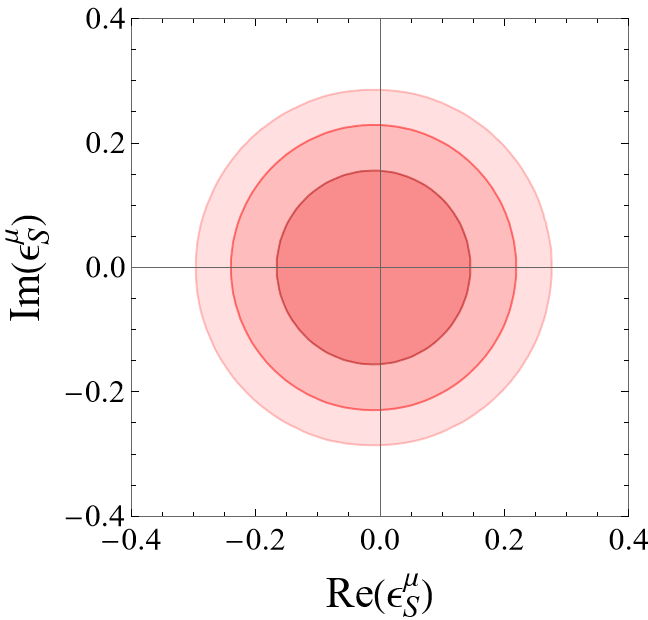}      \hspace*{0.2cm}
\includegraphics[width = 0.23\textwidth]{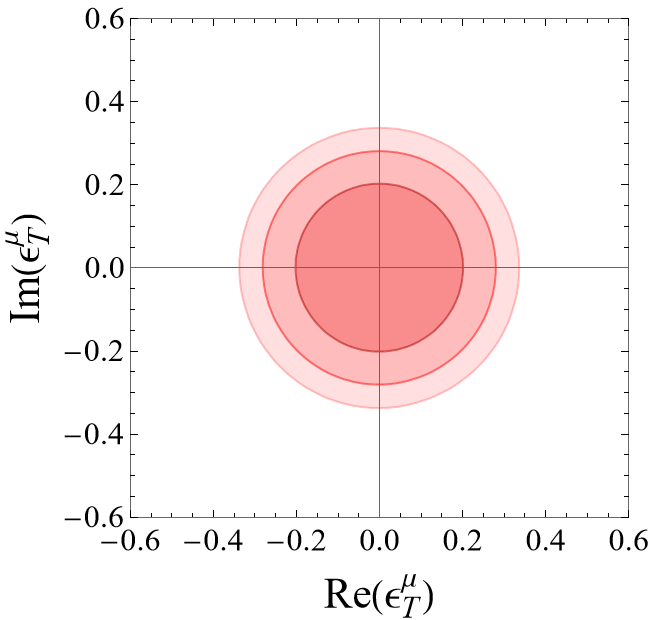}  \\   	\vspace*{0.2cm}	
\includegraphics[width = 0.23\textwidth]{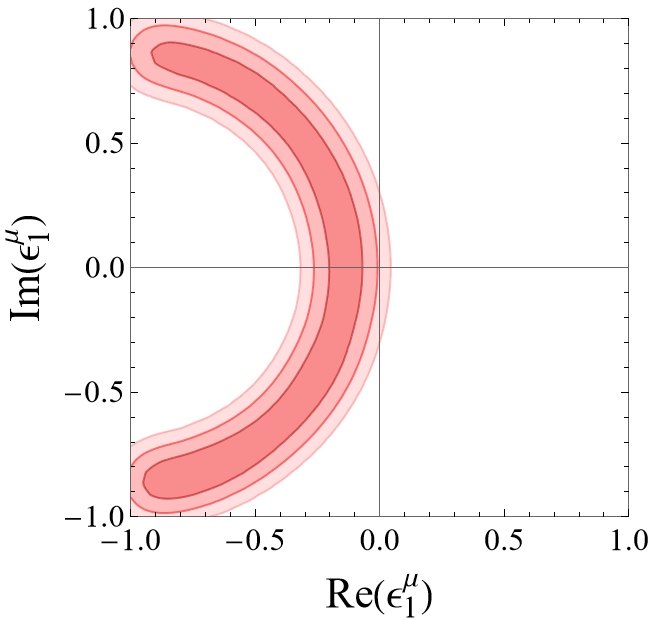} 	    \hspace*{0.2cm}
\includegraphics[width = 0.23\textwidth]{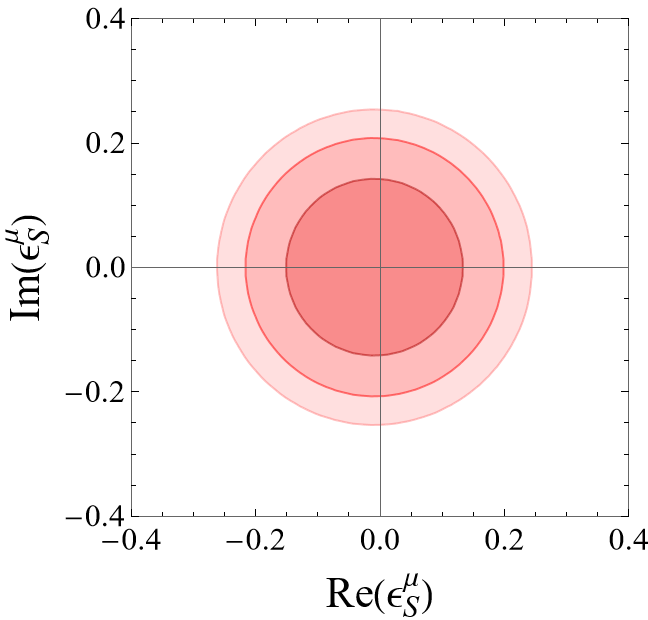}      \hspace*{0.2cm}  
\includegraphics[width = 0.23\textwidth]{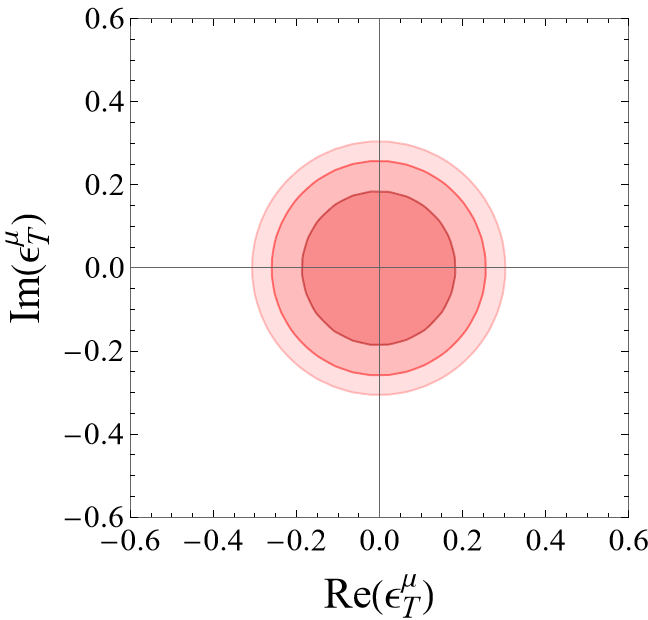}  	
\caption{\baselineskip 10pt  \small   Allowed regions for real and imaginary parts of the NP coefficients $\epsilon_1^\mu$, $\epsilon_S^\mu$ and $\epsilon_T^\mu$, obtained  from the $\bar B^0 \to \pi^+ \mu^- \bar \nu_\mu$ decay mode.  From top to bottom,  the  plots refer to  $|V_{ub}|_{\rm min}$,  $|V_{ub}|_{\rm mid}$, $|V_{ub}|_{\rm max}$, respectively.  The shades of the regions, from the darkest to the lightest one,  correspond to the confidence level $1 \sigma$, $2 \sigma$ and $3 \sigma$.}\label{figpi}
\end{center}
\end{figure*}

Six NP parameters are involved in this decay amplitude, which are the real and imaginary parts of $\epsilon_1^\mu$, $\epsilon_S^\mu$ and $\epsilon_T^\mu$. 
Using the  Monte Carlo method we obtain the regions allowed at   $1\sigma$, $2\sigma$ and $3 \sigma$ C.L. depicted in Fig.~\ref{figpi} for
the   three  benchmark values of $|V_{ub}|$. 
The main constraint is obtained for the coefficient $\epsilon^\mu_S$ of the scalar operator, which becomes more stringent for the largest value of $|V_{ub}|$. Also the coefficient $\epsilon^\mu_T$ of the tensor operator is efficiently constrained, with the same effect induced by the increasing of $|V_{ub}|$.
An interesting feature concerns $\epsilon_1^\mu$, with obtained semi-circular regions  centered at ${\rm Re}(\epsilon_1^\mu)=-1$, ${\rm Im}(\epsilon_1^\mu)=0$:  the $1\sigma$ allowed region favors a nonzero value of ${\rm Re}(\epsilon_1^\mu)$, with the Standard Model point sitting outside this region.

 In the profile-likelihood analysis we set $\Delta \chi^2_{1\sigma}(6)=7.04$. The obtained result  confirms the same preferred NP solution, while yielding  moderately broader confidence regions.

\subsection{$B^- \to \rho^0 \mu^-  {\bar \nu}_\mu$}\label{results-rho}
In this decay mode  the involved NP parameters  are the real and imaginary parts of $\epsilon_1^\mu=\epsilon_V^\mu+\epsilon_R^\mu$, $\epsilon_2^\mu=\epsilon_V^\mu-\epsilon_R^\mu$,  $\epsilon_P^\mu$ and $\epsilon_T^\mu$. The  corresponding  $\Delta \chi^2_{1\sigma}$ is   $\Delta \chi^2_{1\sigma}(8)=9.30$.
Accounting for  the constraint from the purely  leptonic $B^-$ mode we  consider the region  $(-0.06\,\, ,  0.03), (-0.04\,\, ,  0.04)$ for the pair $(\text{Re}(\epsilon_P^{\ell}), \text{Im}(\epsilon_P^{\ell}))$.

\begin{figure*}[t!]
\begin{center}
\includegraphics[width = 0.23\textwidth]{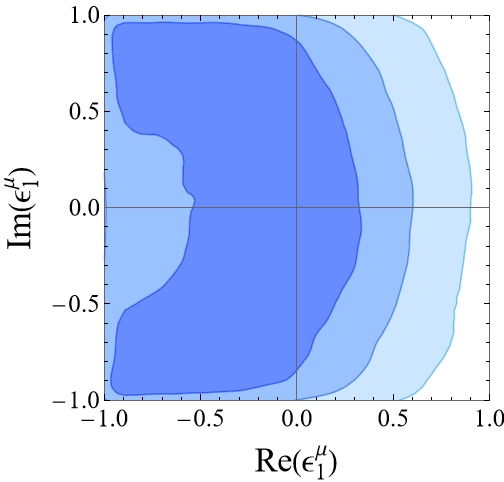} 	\hspace*{0.2cm}
\includegraphics[width = 0.23\textwidth]{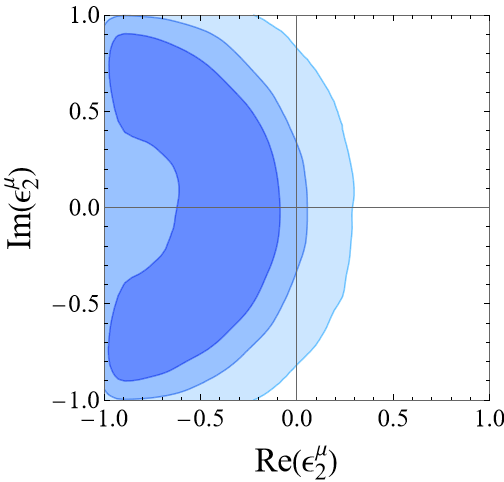} 	\hspace*{0.2cm}
\includegraphics[width = 0.23\textwidth]{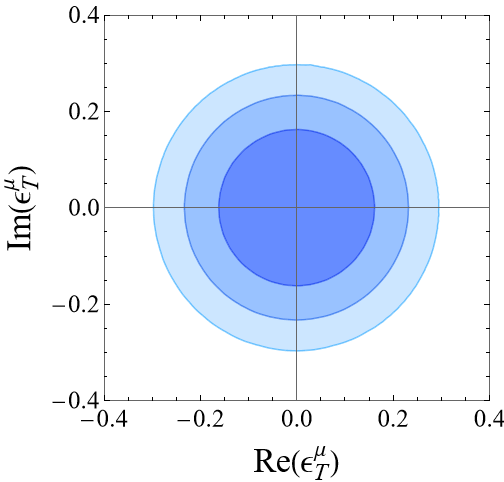}  \\		\vspace*{0.2cm}
\includegraphics[width = 0.23\textwidth]{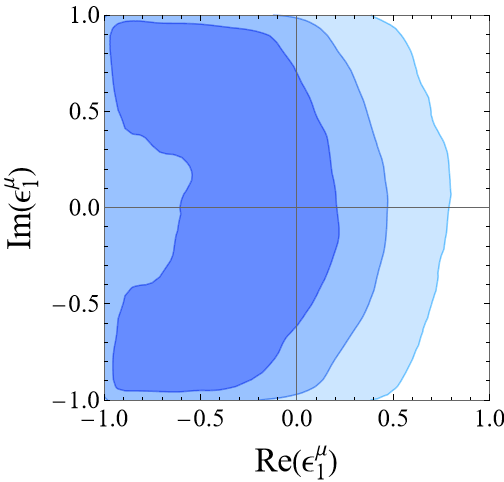} 	\hspace*{0.2cm}
\includegraphics[width = 0.23\textwidth]{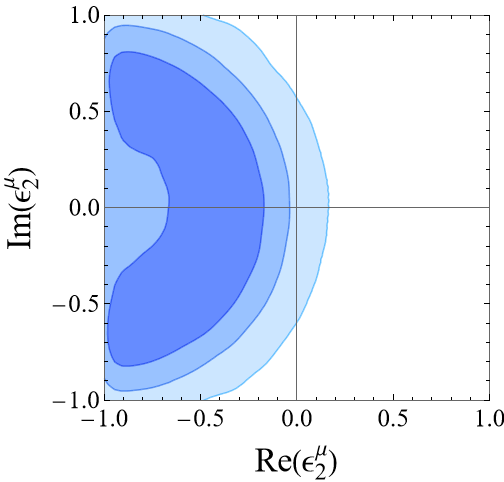} 	\hspace*{0.2cm}
\includegraphics[width = 0.23\textwidth]{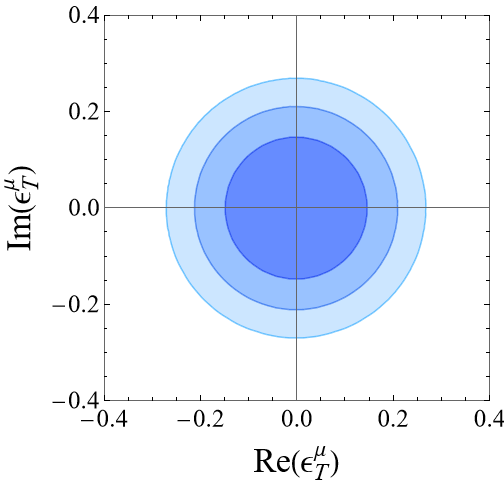}  \\		\vspace*{0.2cm}
\includegraphics[width = 0.23\textwidth]{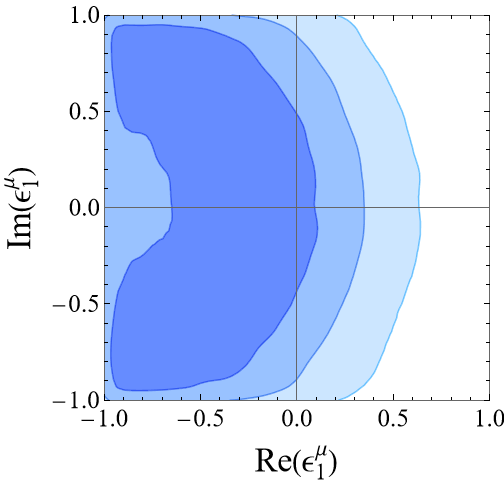} 	\hspace*{0.2cm}
\includegraphics[width = 0.23\textwidth]{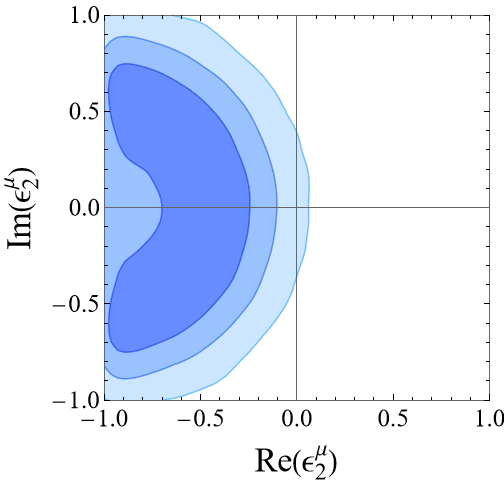} 	\hspace*{0.2cm}
\includegraphics[width = 0.23\textwidth]{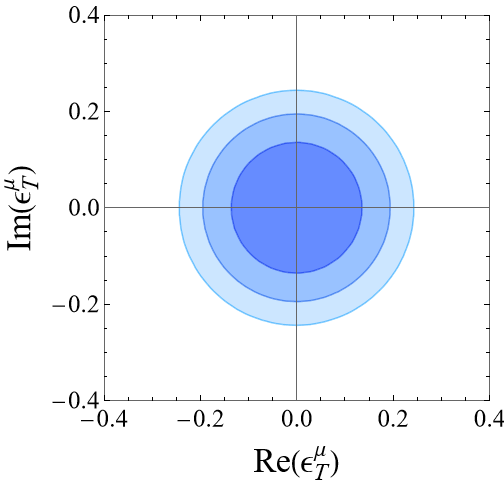}  	
\caption{\baselineskip 10pt  \small  Allowed  regions for the couplings $\epsilon_1^\mu$, $\epsilon_2^\mu$ and $\epsilon_T^\mu$ from the $B^- \to \rho^0 \mu^- \bar \nu_\mu$ mode.  From top to bottom the  plots refer to  $|V_{ub}|_{\rm min}$,  $|V_{ub}|_{\rm mid}$, $|V_{ub}|_{\rm max}$, respectively. The shades of the regions, from the darkest to the lightest one,  correspond to the confidence level $1 \sigma$, $2 \sigma$ and $3 \sigma$. }\label{figrho}
\end{center}
\end{figure*}

Exploiting the Monte Carlo method  we obtain the $1 \sigma$, $2\sigma$, $3 \sigma$ C.L. regions  displayed in Fig.~\ref{figrho}.  We do not show the  $\epsilon_P^\mu$ region since the semileptonic $B^- \to \rho^0$ mode does not  further reduce it   with respect to the purely leptonic channel. A stringent constraint is obtained for the tensor coupling, which is strengthened if the 
value of  $|V_{ub}|$ is increased. On the other hand, the real and imaginary parts of $\epsilon_1^\mu$ and $\epsilon_2^\mu$ are less constrained. The interesting result is that 
 for all  the benchmark values of $|V_{ub}|$ a non-vanishing value of  ${\rm Re}(\epsilon_2^\mu)$ is favoured at $1\sigma$ C.L..  This parameter is compatible with zero only at the $2\sigma$ level  for $|V_{ub}|_{\rm min}$. This  result is not unexpected: actually, in the analysis  performed by the Belle II Collaboration when this channel is separately considered not allowing for  NP contributions, the value  $|V_{ub}|=(3.19 \pm 0.12 \pm 0.17 \pm 0.26) \times 10^{-3}$ has been  obtained (with the errors respectively referring to statistics, systematics and theoretical input)  \cite{Belle-II:2024xwh}. 
 
The profile-likelihood analysis identifies the same preferred solution and also in this case it  yields slightly  broader confidence regions.

 \subsection{Joint analysis of  $B^- \to  \mu^- \bar \nu_\mu$, $\bar B^0 \to \pi^+ \mu^- \bar \nu_\mu$ and  $B^- \to \rho^0 \mu^- \bar \nu_\mu$  }\label{results-combined}
\begin{figure*}[t!]
\begin{center}
\includegraphics[width = 0.19\textwidth]{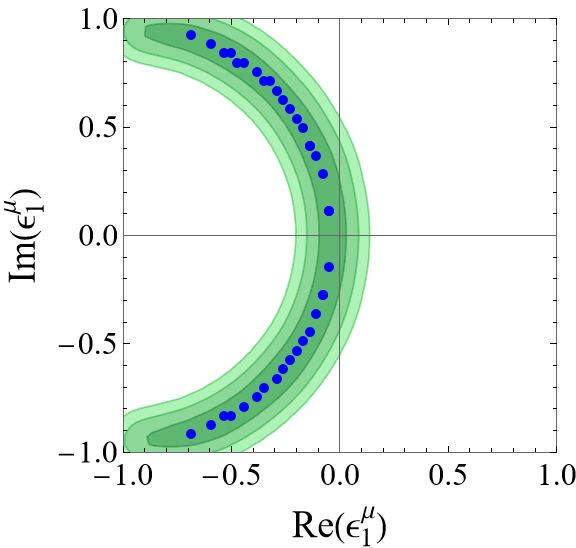} 	
\includegraphics[width = 0.19\textwidth]{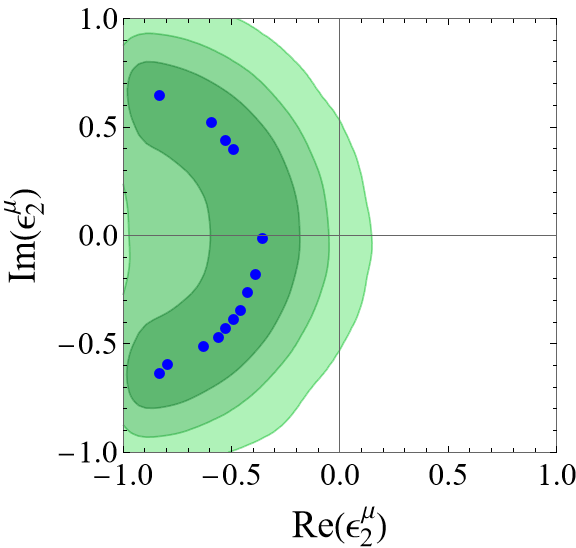}
\includegraphics[width = 0.19\textwidth]{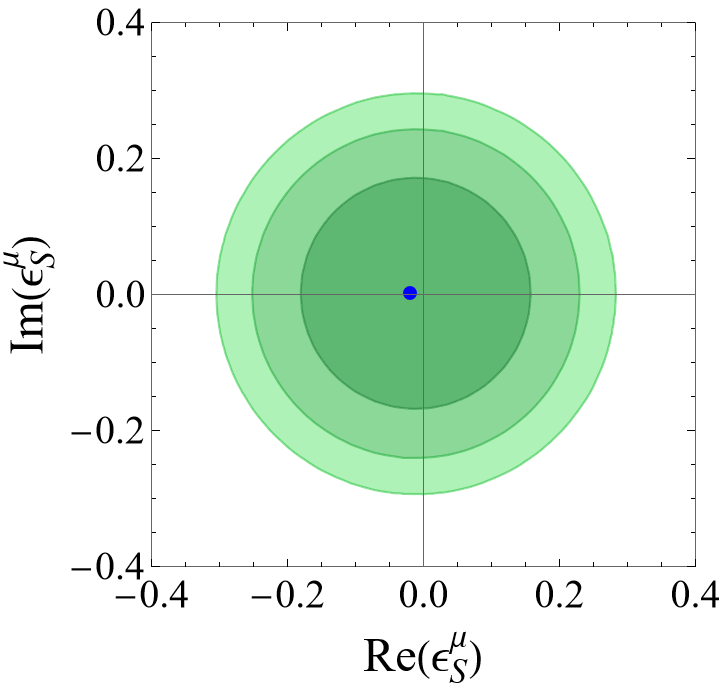}
\includegraphics[width = 0.19\textwidth]{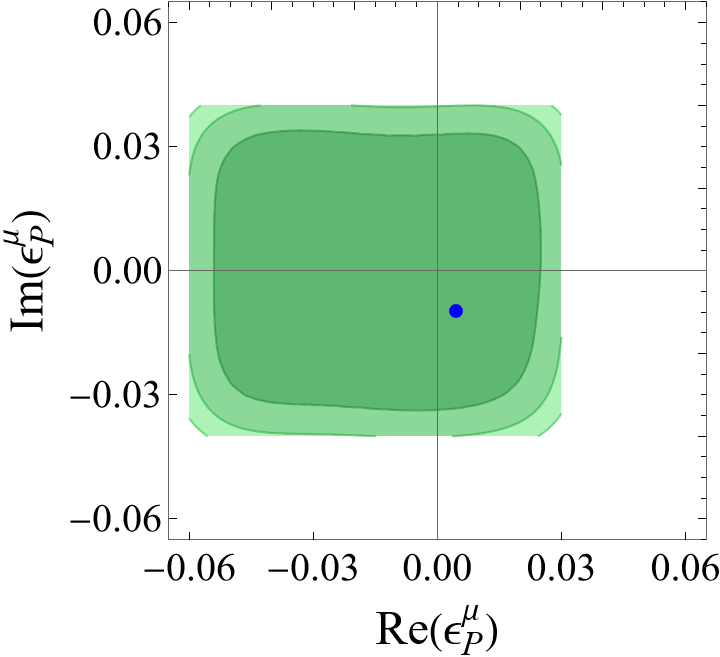}
\includegraphics[width = 0.19\textwidth]{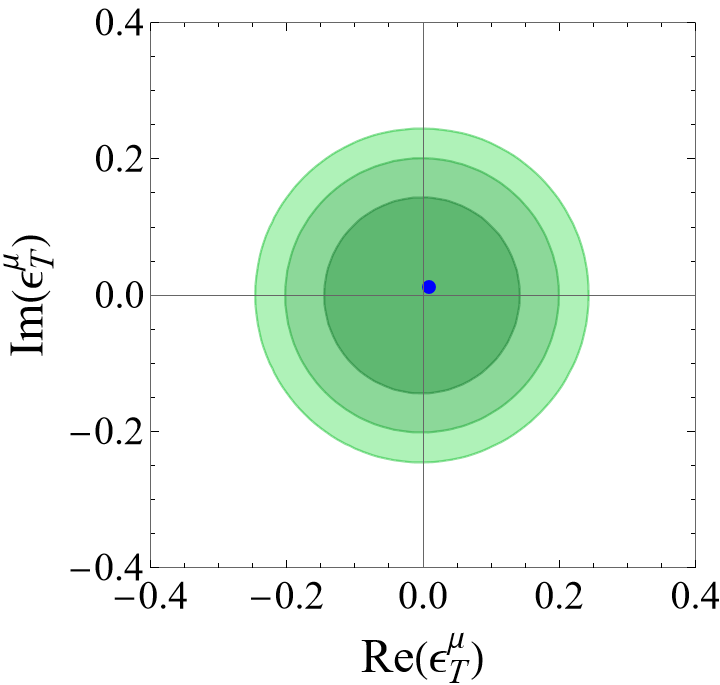}\\
\includegraphics[width = 0.19\textwidth]{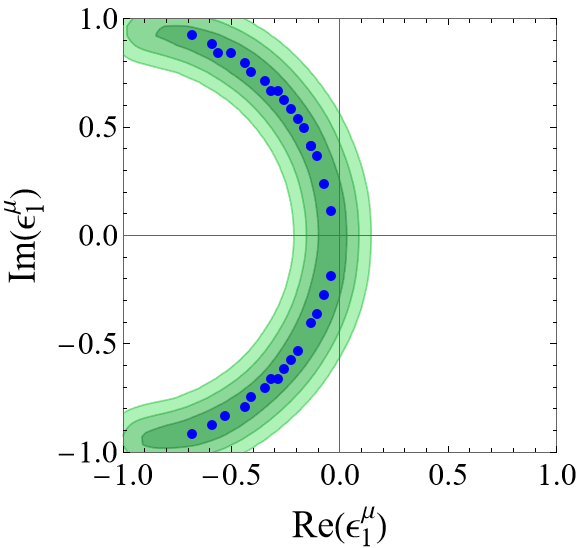} 	
\includegraphics[width = 0.19\textwidth]{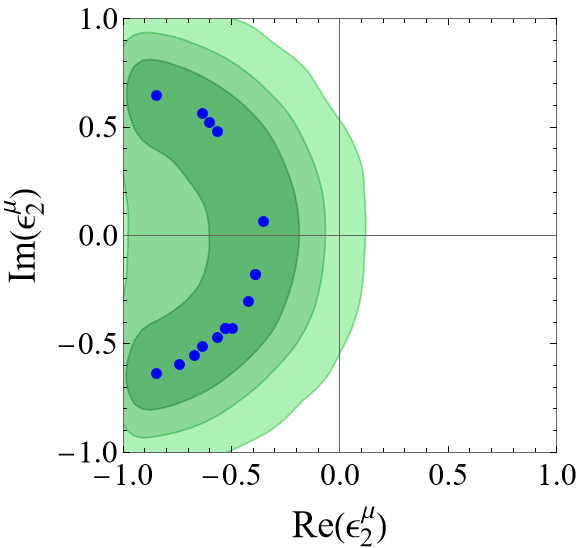}
\includegraphics[width = 0.19\textwidth]{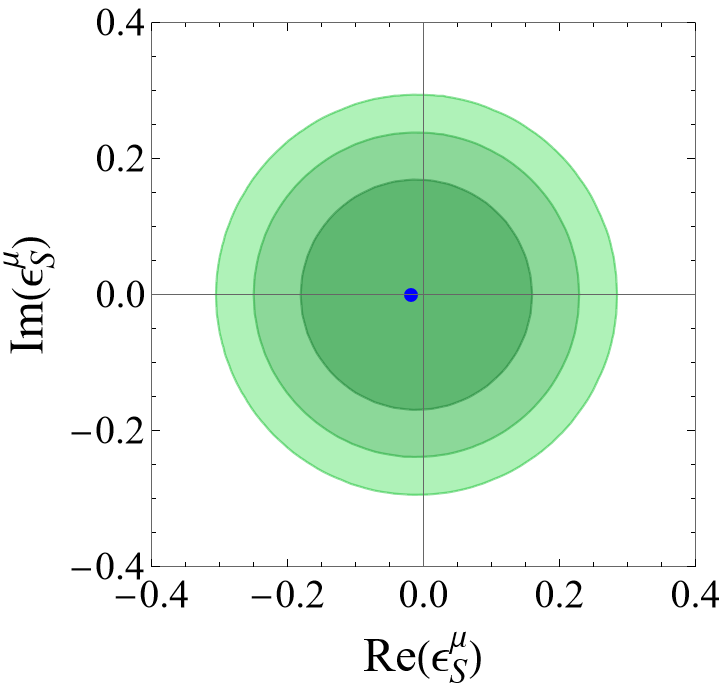}
\includegraphics[width = 0.19\textwidth]{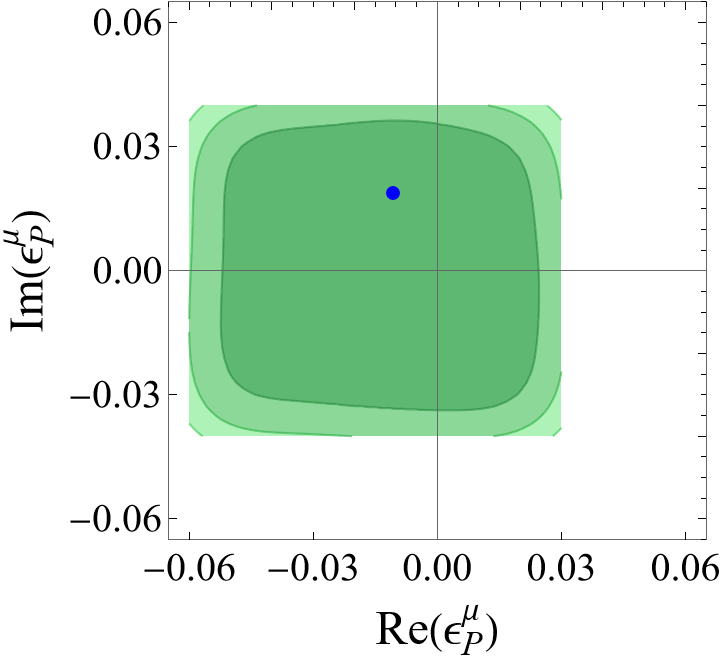}
\includegraphics[width = 0.19\textwidth]{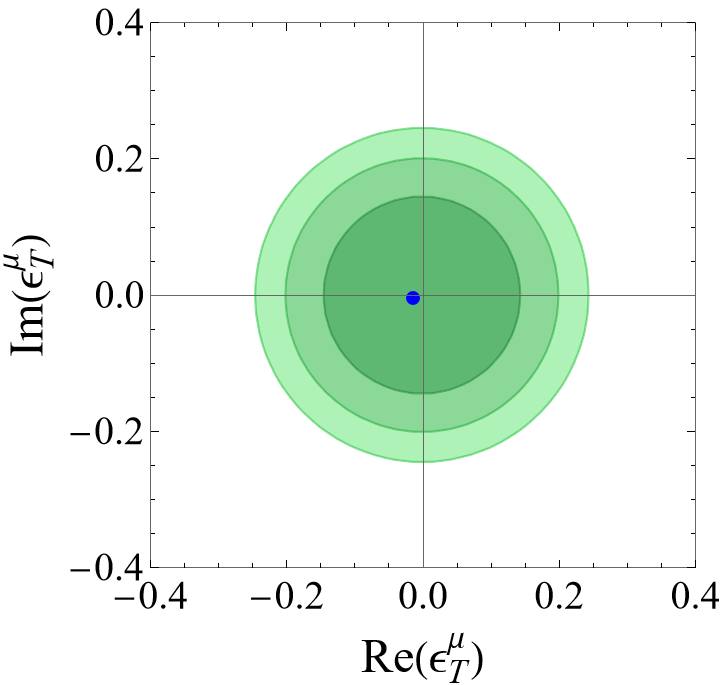}\\
\includegraphics[width = 0.19\textwidth]{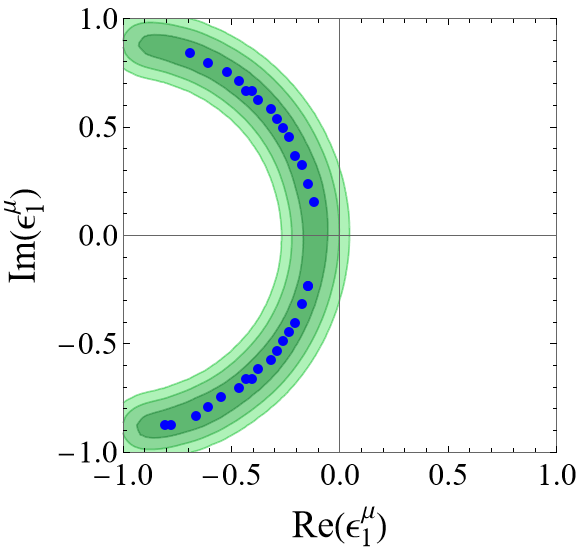} 	
\includegraphics[width = 0.19\textwidth]{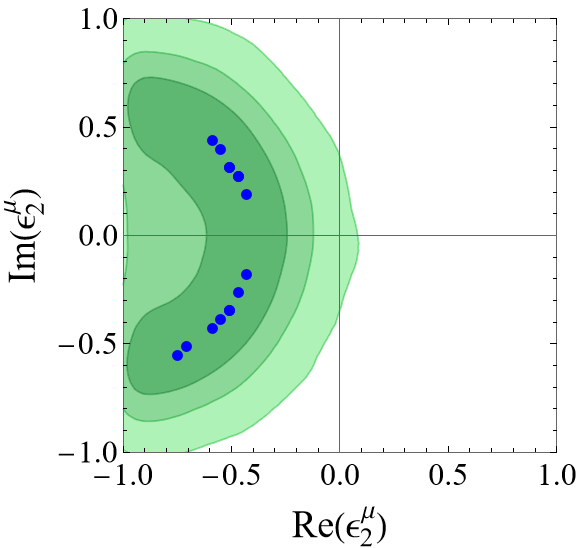}
\includegraphics[width = 0.19\textwidth]{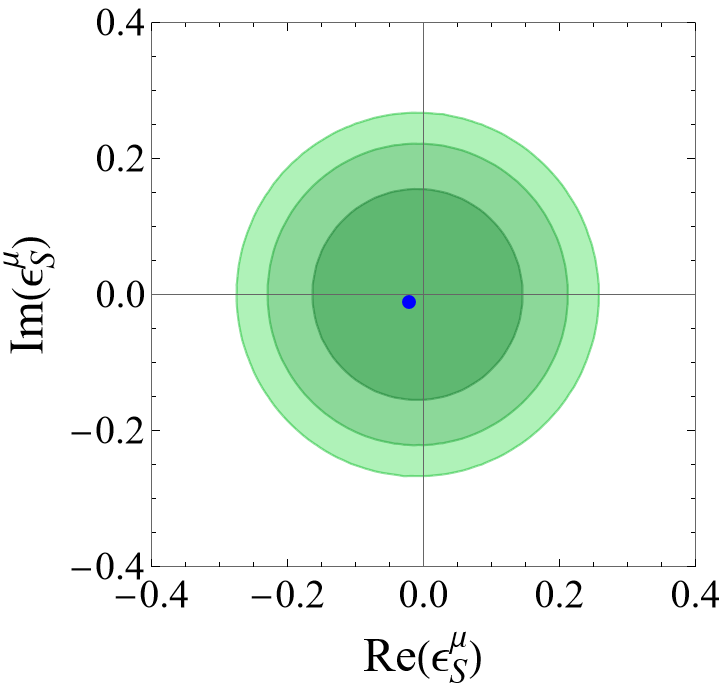}
\includegraphics[width = 0.19\textwidth]{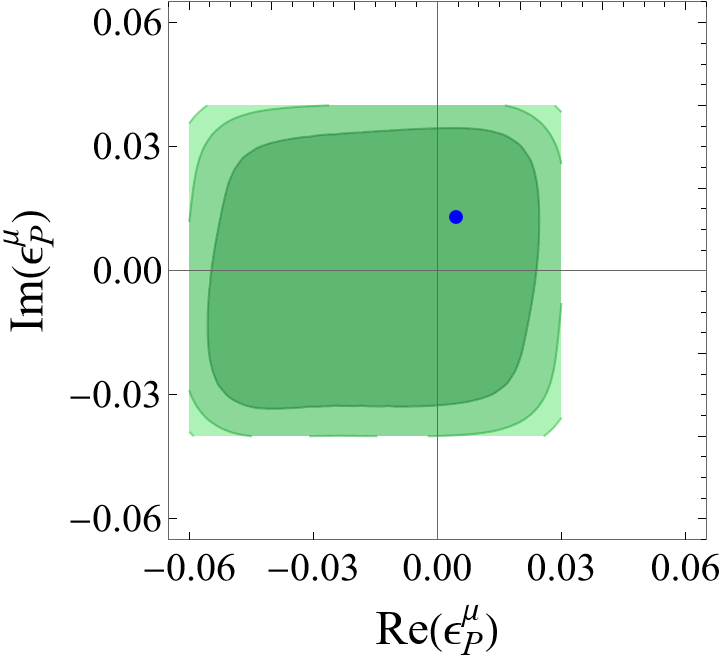}
\includegraphics[width = 0.19\textwidth]{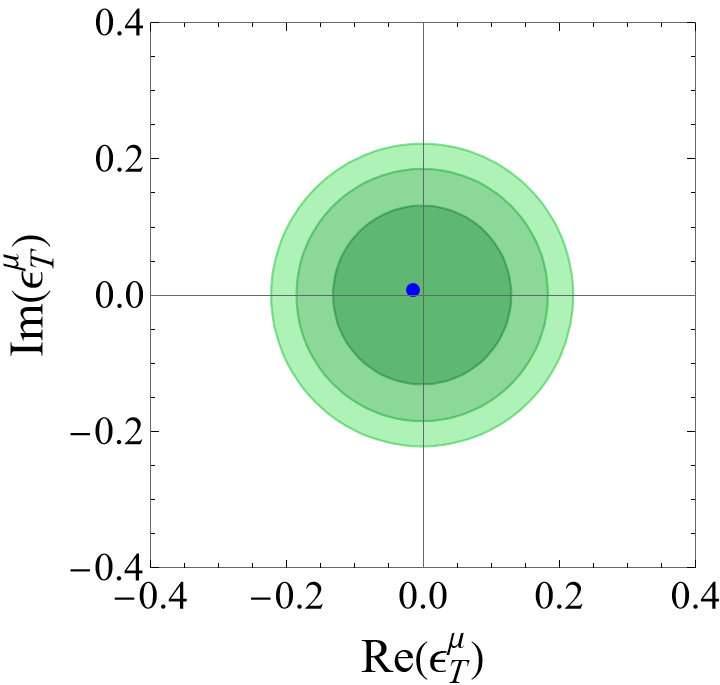}\\
\caption{\baselineskip 10pt  \small  
Allowed  regions for the couplings $\epsilon_1^\mu$, $\epsilon_2^\mu$, $\epsilon_S^\mu$, $\epsilon_P^\mu$  and $\epsilon_T^\mu$ obtained by the combined analysis of the modes $B^- \to  \mu^- \bar \nu_\mu$, $\bar B^0 \to \pi^+ \mu^- \bar \nu_\mu$ and  $B^- \to \rho^0 \mu^- \bar \nu_\mu$.  
  From top to bottom the  plots refer to  $|V_{ub}|_{\rm min}$,  $|V_{ub}|_{\rm mid}$, $|V_{ub}|_{\rm max}$, respectively.  The  shades of green, from the darkest to the lightest one, indicate
  the  parameter regions allowed at $1 \sigma$, $2 \sigma$ and $3 \sigma$. The dots correspond to the best-fit values, see the text for the  discussion in the case of $\epsilon_1^\mu$ and 
  $\epsilon_2^\mu$. }\label{figAll}
\end{center}
\end{figure*}
After separately considering the purely leptonic and the semileptonic  $\pi^+$ and $\rho^0$ channels, we  now study how the full set of  decay  modes  provides constraints for the NP parameters in the low-energy Hamiltonian \eqref{heff}. 

To apply  the Monte Carlo method, we construct the total likelihood, and then we determine the  best-fit points and  the allowed regions in the parameter space.
The results  for the best-fit values are displayed in Table \ref{tabrho}.  A small preference is found for the $|V_{ub}|_{\rm mid}$ case, as one can infer considering the value of the log-likelihood  for the three benchmark values.

The results for the  parameter regions are shown in Fig.~\ref{figAll} for the three choices of $|V_{ub}|$  at  $1 \sigma$, $2\sigma$ and $3 \sigma$ C.L. 
 For each pair of the real and imaginary part of couplings, the best-fit estimate -- the global maximum of the marginalized posterior $p\bigl(\text{Re}(\epsilon_{j}^{\mu}),\text{Im}(\epsilon_{j}^{\mu})|\,\text{data}\bigr)$ -- is determined from  the kernel approximation $k\bigl(\text{Re}(\epsilon_{j}^{\mu}),\text{Im}(\epsilon_{j}^{\mu})\bigr)$. However, the topology of the probability regions in the complex plane of some cases, namely  $\epsilon_{1}^\mu$ and $\epsilon_{2}^\mu$, suggests the presence of a  direction with degenerate maxima. To investigate this, we implemented a grid-seeded continuous search for local maxima of $k$.\footnote{This is performed by evaluating  the function $k$ on a $200\times 200$ grid in the complex plane; the found local maxima are then used as starting points for the \texttt{FindMaximum} routine in Wolfram \textit{Mathematica}.} The posterior densities for $\epsilon_1^\mu$ and $\epsilon_2^\mu$ are found to possess multiple degenerate maxima distributed along a semicircular ridge centered at $(-1,0)$ and radius $r < 1$. The fluctuations in peak heights 
are consistent with the finite resolution of the Monte Carlo sampling and of the kernel density estimation, an indication that the local maxima are statistically equivalent. The conclusion is that, while the coefficients $\epsilon_S^\mu$, $\epsilon_P^\mu$, and $\epsilon_T^\mu$ possess  isolated  global maxima of the probability function, there is a  degeneracy of maxima in such a function for $\epsilon_1^\mu$ and $\epsilon_2^\mu$. This is illustrated in Fig.~\ref{figAll}, in which the position of the maximum probability is indicated  by dark dots. This  also explains the change of sign of the coordinates of some best-fit points in Table \ref{tabrho} varying $|V_{ub}|$.
The remarkable result of the combined analysis is that ${\rm Re}(\epsilon_2^\mu)=0$ is disfavoured at $2 \sigma$ level for all values of $|V_{ub}|$, and that
 ${\rm Re}(\epsilon_1^\mu)=0$ is disfavoured at  $1 \sigma$ for $|V_{ub}|_{\rm max}$.

 \subsection{Global fit and determination of $ |V_{ub}|$ }\label{all-results}
We finally perform a global fit in which also $|V_{ub}|$ is included in the set of parameters, together with the NP couplings.  The result 
\be
|V_{ub}| = (3.79\pm0.17) \times 10^{-3} \,\,,  \label{Vubfit}
\ee
is slightly smaller than the outcome  \eqref{VubSM} obtained in the SM.
 The selected regions of the NP couplings are similar to the case $|V_{ub}|_{\rm mid}$ in Fig.~\ref{figAll}

\section{Discussion and outlook}\label{conclusion}

We  summarize our findings.
We have investigated the implications of the recent Belle II measurements  of the binned $q^2$ distributions of  the semileptonic $\bar B^0 \to \pi^+$ and  $ B^- \to \rho^0$ muon  decays, together with the current upper limit on the purely muonic $B^-$ decay mode, for the most general low-energy Hamiltonian describing $b \to u \, \mu \bar \nu_\mu$ with left-handed neutrinos. 
 The analysis has been performed in a model-independent framework, allowing all dimension-six Wilson coefficients to vary simultaneously. The three decay modes provide complementary information on the different operator structures. 
 
\begin{itemize}
\item  For the   purely leptonic $B^- \to \mu^- {\bar \nu}_\mu$  mode, the pseudoscalar NP operator 
 removes the SM chiral suppression.  As a consequence, the upper bound on the decay width  provides the  strongest constraint on   $\epsilon_P^\mu$.
 \item
The $\bar B^0 \to \pi^+ \mu^- {\bar \nu}_\mu$  decay amplitude receives contributions from vector, scalar, and tensor operators. The $q^2$ distribution depends on specific linear combinations of the left- and right-handed vector couplings, which interfere with the SM contribution  modifiyng  the normalization and the shape of the 
spectrum. The scalar contribution governed by $\epsilon_S^\ell$ increases with $q^2$. The tensor operator controlled by  $\epsilon_T^\mu$  induces a distinct kinematic dependence,  and can be  probed through a distortion of the spectrum.

\item For   $B^- \to \rho^0 \mu^- {\bar \nu}_\mu$,  the rich helicity structure of the decay induces the sensitivity to a larger set of operators. The vector, pseudoscalar, and tensor couplings enter the helicity amplitudes, resulting in a more involved  dependence of the spectrum on the Wilson coefficients. The tensor operator contribution can play a   prominent role due to the coupling to transverse  polarization states of the $\rho$ meson.
\end{itemize}

Their combined analysis therefore leads to stronger constraints than those obtained from any individual decay channel. 
To assess the robustness of the extracted parameter regions, we have employed two independent statistical approaches. The  MCMC analysis has been used to determine the highest-posterior-density regions, while an independent profile-likelihood analysis, accelerated through a Random-Forest emulator, has been performed as a cross-check. The two approaches lead to  consistent confidence regions and compatible best-fit points, with  the profile-likelihood regions  slightly broader than the corresponding HPD  ranges.

The combined analysis favours non-vanishing values of $\mathrm{Re}(\epsilon_2^\mu)$ for all benchmark values of $|V_{ub}|$, with $\mathrm{Re}(\epsilon_2^\mu)=0$ lying outside the $2\sigma$ allowed region. Moreover, for the largest benchmark value of $|V_{ub}|$, the point $\mathrm{Re}(\epsilon_1^\mu)=0$ is also excluded at the $1\sigma$ level. While these results cannot be interpreted as evidence for physics beyond the Standard Model, they indicate regions of the parameter space that deserve further scrutiny as more precise measurements become available.
Finally, we have determined $|V_{ub}|$ allowing simultaneously for generic new-physics contributions. The extracted value is compatible with the Standard Model determination, demonstrating that the present experimental information already permits a meaningful simultaneous fit of CKM and new-physics parameters within the effective-field-theory framework.

The prospects for tightening these constraints are promising. The growing Belle II data sample, combined with measurements of fully differential $B^-\to\rho^0(\pi^+\pi^-)\mu^-\bar\nu_\mu$ distributions and further progress in the determination of the relevant hadronic form factors, will significantly enhance the sensitivity to the Lorentz structure of the underlying interactions \cite{Colangelo:2019axi}. Future global analyses of $b\to u$ semileptonic transitions will therefore provide increasingly stringent tests of the Standard Model and powerful probes of potential new physics effects.

\section*{Acknowledgements}
This work has  been carried out within the INFN project (Iniziativa Specifica)  SPIF.

\begin{widetext}
 \begin{table*}[h!]
\begin{tabular}{ c c c c c c c c  c c c c} 
\hline \hline
$|V_{ub}| \times 10^3$ &  ${\ln }{\cal L} _{\rm max}$ & ${\rm Re}(\epsilon_1^\mu)$ & ${\rm Im}(\epsilon_1^\mu)$& ${\rm Re}(\epsilon_2^\mu)$ & ${\rm Im}(\epsilon_2^\mu)$
& ${\rm Re}(\epsilon_S^\mu)$ & ${\rm Im}(\epsilon_S^\mu)$& ${\rm Re}(\epsilon_P^\mu)$ & ${\rm Im}(\epsilon_P^\mu)$& ${\rm Re}(\epsilon_T^\mu)$ & ${\rm Im}(\epsilon_T^\mu)$\\ \hline 
 $3.51$ & 288.790 & -0.190	& 0.771	 &-0.856	 &-0.715	  &-0.057   &0.020  &-0.046 &  0.031 &  -0.090  & -0.082  \\
$3.85$ & 288.789& -0.582	&-0.937	& -0.884	 & 0.705	 &  0.017  & 0.072 & -0.036 &  -0.026 &  -0.009  &  0.009   \\
$4.19$ & \,\,288.798 \,\,&-0.738 &0.896	&  -0.905	& -0.591	&  -0.005 &  -0.036 & -0.014 &  0.006  & 0.020 &  -0.108 \\
\hline \hline
\end{tabular} 
 \caption{\baselineskip 10pt  \small  Best-fit point in the space of the new physics couplings obtained from the combined analysis of   $B^- \to \mu^- {\bar \nu}_\mu$,  $\bar B^0 \to \pi^+ \mu^- {\bar \nu}_\mu$ and  $B^- \to \rho^0 \mu^- {\bar \nu}_\mu$   for  the three benchmark values of $|V_{ub}|$. The corresponding maximum value of the log-likelihood is indicated. To change of sign in some results observed when $|V_{ub}|$ is changed is discussed in the text. }
\label{tabrho} 
\end{table*}
\end{widetext}

\begin{widetext}
\appendix
\section{Quark current matrix elements}\label{app-ff}

We use the standard parametrization of the $B \to \pi$ and $B \to \rho$ matrix elements of quark currents in terms of form factors.

\qquad $\bar B^0 \to \pi^+$:
\bea
\langle \pi(p^\prime)| {\bar u} \gamma_\mu b| {\bar B}(p) \rangle &=& f_+^{B \to \pi}(q^2) \Big[p_\mu+p_\mu^\prime  - \frac{m_B^2-m_\pi^2}{q^2} q_\mu\Big]
+f_0^{B \to \pi}(q^2)\frac{m_B^2-m_\pi^2}{q^2} q_\mu  \label{FF-pi1} \\ 
\langle \pi(p^\prime)| {\bar u} b| {\bar B}(p) \rangle & =& f_S^{B \to \pi}(q^2)  \label{FF-pi2}  \\
\langle \pi(p^\prime)| {\bar u} \sigma_{\mu \nu }b| {\bar B}(p) \rangle &=& -i \frac{2 f_T^{B \to \pi}(q^2)}{m_B+m_\pi} \big[p_\mu p_\nu^\prime-p_\nu p^\prime_\mu \big] \label{FF-pi3}  \\
\langle \pi(p^\prime)| {\bar u} \sigma_{\mu \nu }\gamma_5 b| {\bar B}(p) \rangle &=&- \frac{2 f_T^{B \to \pi}(q^2)}{m_B+m_\pi} \epsilon_{\mu \nu \alpha \beta} \, p^\alpha p^{\prime \beta} \,\,\, , \label{FF-pi4} 
\eea
with $\epsilon^{0123}=+1$ and  $f_+^{B \to \pi}(0)=  f_0^{B \to \pi}(0)$.  $f_S$ and $f_0$ are related:  $f_S^{B \to \pi}(q^2)=\displaystyle \frac{m_B^2-m_\pi^2}{m_b-m_u}f_0^{B \to \pi}(q^2)$.

\qquad  $\bar B^0 \to \rho^+$:
\bea
\langle \rho(p^\prime,\epsilon)|{\bar u} \gamma_\mu b | {\bar B}(p) \rangle &=&
- {2 V^{B \to \rho} (q^2) \over m_B+m_\rho} i \epsilon_{\mu \nu \alpha \beta} \epsilon^{*\nu}  p^\alpha p^{\prime \beta}   \label{FF-rho0}  \\
\langle \rho(p^\prime,\epsilon)|{\bar u} \gamma_\mu \gamma_5 b | {\bar B}(p) \rangle &=&
\Big\{ (m_B+m_\rho) \left[ \epsilon^*_\mu -{(\epsilon^* \cdot q) \over q^2} q_\mu \right] A_1^{B \to \rho}(q^2) \nn\\
&+& {(\epsilon^* \cdot q) \over  m_B+m_\rho} \left[ (p+p^\prime)_\mu -{m_B^2-m_\rho^2 \over q^2} q_\mu \right] A_2^{B \to \rho}(q^2) 
-(\epsilon^* \cdot q){2 m_\rho \over q^2} q_\mu A_0^{B \to \rho}(q^2) \Big\} \qquad  \label{FF-rho1} 
\eea
with   $\displaystyle A_0^{B \to \rho}(0)= \frac{m_B + m_\rho}{2 m_\rho} A_1^{B \to \rho}(0) -  \frac{m_B - m_\rho}{2 m_\rho}  A_2^{B \to \rho}(0)$,
\bea
\langle \rho(p^\prime,\epsilon)|{\bar u} \gamma_5 b| {\bar B}(p) \rangle &=&-\frac{2 m_\rho}{m_b+m_u} (\epsilon^* \cdot q) A_0^{B \to \rho}(q^2)
\label{FF-rho2} \\
\langle \rho(p^\prime,\epsilon)|{\bar u} \sigma_{\mu \nu} b| {\bar B}(p) \rangle& =&
T_0^{B \to \rho}(q^2) {\epsilon^* \cdot q \over (m_B+ m_\rho)^2} \epsilon_{\mu \nu \alpha \beta} p^\alpha p^{\prime \beta}\nn \\
&+&
T_1^{B \to \rho}(q^2) \epsilon_{\mu \nu \alpha \beta} p^\alpha \epsilon^{*\beta}+ T_2^{B \to \rho}(q^2) \epsilon_{\mu \nu \alpha \beta} p^{\prime \alpha} \epsilon^{*\beta} \label{FF-rho3}\\
\langle \rho(p^\prime,\epsilon)|{\bar u} \sigma_{\mu \nu}\gamma_5 b| {\bar B}(p) \rangle &=&
i\, T_0^{B \to \rho}(q^2) {\epsilon^* \cdot q \over (m_B+ m_\rho)^2} (p_\mu p^\prime_\nu-p_\nu p^\prime_\mu) \nn \\
& +&i\,
T_1^{B \to \rho}(q^2) (p_\mu \epsilon_\nu^*-\epsilon_\mu^* p_\nu)+i\,T_2^{B \to \rho}(q^2)(p^\prime_\mu \epsilon_\nu^*-\epsilon_\mu^* p^\prime_\nu) \,\,. \,\,\, \label{FF-rho4}
\eea
%
\section{Angular coefficient functions}\label{appcoeff}

The fully differential $B \to \rho (\pi \pi) \ell \bar \nu_\ell$  decay distribution in terms of angular coefficient functions is given in \cite{Colangelo:2019axi}.  The coefficient functions $I_{1s}, I_{1c}, I_{2s}, I_{2c}$ determining the $q^2$ distribution  are written in the form
\bea
I_i &=& |1+\epsilon_V|^2 \,I_i^{SM}+|\epsilon_P|^2I_i^{NP,P}+|\epsilon_T|^2I_i^{NP,T} 
+ 2 \, {\rm Re}\left[\epsilon_P(1+\epsilon_V^* )\right] I_i^{INT,P} 
+2 \, {\rm Re}\left[\epsilon_T(1+\epsilon_V^* )\right] I_i^{INT,T} \nn \\
&+&2 \, {\rm Re}\left[\epsilon_P \epsilon_T^* \right] I_i^{INT,PT}  \quad , \label{eq:Iang}
\eea
with $i=1s,1c,2s,2c$.
The expressions of the  functions in \eqref{eq:Iang} are given  in Table  \ref{tab:rhoSM}.
They   involve  the helicity amplitudes in SM \cite{Colangelo:2018cnj}

\bea
H_0 &=&\frac{(m_B+m_\rho)^2(m_B^2-m_\rho^2-q^2) A_1^{B \to \rho} (q^2)-\lambda(m_B^2,\,m_\rho^2,\,q^2) A_2^{B \to \rho} (q^2)}{2m_\rho(m_B+m_\rho) \sqrt{q^2}} \label{Hamprho0}\\
H_\pm&=& \frac{(m_B+m_\rho)^2 A_1^{B \to \rho} (q^2)\mp\sqrt{\lambda(m_B^2,\,m_\rho^2,\,q^2)}V^{B \to \rho} (q^2)}{m_B+m_\rho}  \label{Hamprho1}\\
H_t&=& -\frac{\sqrt{\lambda(m_B^2,\,m_\rho^2,\,q^2)}}{\sqrt{q^2}} \,A_0^{B \to \rho} (q^2) \,\,\, \label{Hamprho2} 
\eea
and
\bea
H_+^{NP} &=& \frac{1}{\sqrt{q^2}}\left\{\left[m_B^2-m_\rho^2+\lambda^{1/2} (m_B^2,m_\rho^2,q^2) \right](T_1^{B \to \rho}+ T_2^{B \to \rho})+q^2(T_1^{B \to \rho}- T_2^{B \to \rho})\right\}  \\
H_-^{NP} &=& \frac{1}{\sqrt{q^2}}\left\{\left[m_B^2-m_\rho^2-\lambda^{1/2} (m_B^2,m_\rho^2,q^2) \right](T_1^{B \to \rho}+ T_2^{B \to \rho})+q^2(T_1^{B \to \rho}-T_2^{B \to \rho})\right\} \hspace*{1cm} \\
H_L^{NP}&=&4\Big\{
\frac{\lambda (m_B^2,m_\rho^2,q^2)}{m_\rho(m_B+m_\rho)^2} \, T_0^{B \to \rho}+2\frac{m_B^2+m_\rho^2-q^2}{m_\rho}\, T_1^{B \to \rho}
+4m_\rho\, T_2^{B \to \rho} \Big\} \,\, 
\eea
obtained from the tensor operator in \eqref{heff}.
\end{widetext}

\begin{table*}[t]
\centering
\begin{tabular}{c c c c }
\hline \hline
\noalign{\smallskip}
$i$ & $I_i^{\sm}$ &$I_i^{\np,P}$ & $I_i^{\inter,P}$ \\
\noalign{\smallskip}
\hline
\noalign{\smallskip}
$I_{1s}$ \quad& $\frac{1}{2}(H_+^2 + H_-^2)(m_{\ell}^2 + 3 q^2)$ \quad&$0$ & $0$  \\
$I_{1c}$ \quad& $4 m_{\ell}^2 H_t^2 + 2 H_0^2 (m_{\ell}^2 + q^2)$ \quad &$4 H_t^2 \frac{q^4}{(m_b + m_u)^2}$ \quad & $4 H_t^2 \frac{m_{\ell} q^2}{m_b + m_u}$ \\
$I_{2s}$ \quad & $- \frac{1}{2}(H_+^2 + H_-^2)(m_{\ell}^2 - q^2)$ \quad &$0$ & $0$ \\
$I_{2c}$ \quad & $2 H_0^2 (m_{\ell}^2 - q^2)$ \quad &$0$ & $0$  \\
\noalign{\smallskip}
\hline 
\hline 
\noalign{\smallskip}
$i$ & $I_i^{\np,T}$ & $I_i^{\inter, T}$ \\
\noalign{\smallskip}
\hline
\noalign{\smallskip}
$I_{1s}$ & $2 \big[(H_+^\np)^2+(H_-^\np)^2 \big](3 m_\ell^2 + q^2)$ \quad & \quad $-4 ( H_+^\np H_+ + H_-^\np H_- ) m_\ell \sqrt{q^2}$ \\
$I_{1c}$ & $\frac{1}{8} (H_L^\np)^2 (m_\ell^2 + q^2)$ & $- H_L^\np H_0 m_\ell \sqrt{q^2}$ \\
$I_{2s}$ & $2 \big [(H_+^\np)^2+(H_-^\np)^2 \big](m_\ell^2 - q^2)$ & $0$ \\
$I_{2c}$ & $\frac{1}{8} (H_L^\np)^2 (q^2 - m_\ell^2)$ & $0$ \\
\noalign{\smallskip}
\hline \hline
\end{tabular}
\caption{\small Coefficient functions  in \eqref{eq:Iang}. $I_i^{SM}$ are involved in  SM;  $I_i^{\np,P}$  are the terms from  the pseudoscalar P operator;   $I_i^{\inter,P}$  derive  from the interference terms   between the SM and  P  operators;  
$I_i^{\np,T}$  are the term from the tensor T operator;  $I_i^{\inter, T}$ derive from the  interference terms between the SM and   T operators. All the interference terms  $I_i^{INT, PT}$  between P and  tensor T operators vanish. }\label{tab:rhoSM}
\end{table*}

\bibliographystyle{apsrev4-1}
\bibliography{refDFLP}

@article{Cabibbo:1963yz,
    author = "Cabibbo, Nicola",
    title = "{Unitary Symmetry and Leptonic Decays}",
    doi = "10.1103/PhysRevLett.10.531",
    journal = "Phys. Rev. Lett.",
    volume = "10",
    pages = "531--533",
    year = "1963"
}

@article{Kobayashi:1973fv,
    author = "Kobayashi, Makoto and Maskawa, Toshihide",
    title = "{CP Violation in the Renormalizable Theory of Weak Interaction}",
    reportNumber = "KUNS-242",
    doi = "10.1143/PTP.49.652",
    journal = "Prog. Theor. Phys.",
    volume = "49",
    pages = "652--657",
    year = "1973"
}

@article{Belle:2010hep,
    author = "Ha, H. and others",
    collaboration = "Belle",
    title = "{Measurement of the decay $B^0\to\pi^-\ell^+\nu$ and determination of $|V_{ub}|$}",
    eprint = "1012.0090",
    archivePrefix = "arXiv",
    primaryClass = "hep-ex",
    reportNumber = "BELLE-PREPRINT-2010-22, KEK-PREPRINT-2010-37",
    doi = "10.1103/PhysRevD.83.071101",
    journal = "Phys. Rev. D",
    volume = "83",
    pages = "071101",
    year = "2011"
}

@article{BaBar:2010efp,
    author = "del Amo Sanchez, P. and others",
    collaboration = "BaBar",
    title = "{Study of $B \to \pi \ell \nu$ and $B \to \rho \ell \nu$ Decays and Determination of $|V_{ub}|$}",
    eprint = "1005.3288",
    archivePrefix = "arXiv",
    primaryClass = "hep-ex",
    reportNumber = "SLAC-PUB-14106, BABAR-PUB-09-037",
    doi = "10.1103/PhysRevD.83.032007",
    journal = "Phys. Rev. D",
    volume = "83",
    pages = "032007",
    year = "2011"
}

@article{BaBar:2012thb,
    author = "Lees, J. P. and others",
    collaboration = "BaBar",
    title = "{Branching fraction and form-factor shape measurements of exclusive charmless semileptonic B decays, and determination of $|V_{ub}|$}",
    eprint = "1208.1253",
    archivePrefix = "arXiv",
    primaryClass = "hep-ex",
    reportNumber = "BABAR-PUB12-015, SLAC-PUB-15208",
    doi = "10.1103/PhysRevD.86.092004",
    journal = "Phys. Rev. D",
    volume = "86",
    pages = "092004",
    year = "2012"
}

@article{Belle:2013hlo,
    author = "Sibidanov, A. and others",
    collaboration = "Belle",
    title = "{Study of Exclusive $B \to X_u \ell \nu$ Decays and Extraction of $\|V_{ub}\|$ using Full Reconstruction Tagging at the Belle Experiment}",
    eprint = "1306.2781",
    archivePrefix = "arXiv",
    primaryClass = "hep-ex",
    reportNumber = "BELLE-PREPRINT-2013-9, KEK-PREPRINT-2013-8",
    doi = "10.1103/PhysRevD.88.032005",
    journal = "Phys. Rev. D",
    volume = "88",
    number = "3",
    pages = "032005",
    year = "2013"
}

@article{CLEO:1999mif,
    author = "Behrens, B. H. and others",
    collaboration = "CLEO",
    title = "{Measurement of $B \to \rho$ lepton neutrino decay and $|V_{ub}|$}",
    eprint = "hep-ex/9905056",
    archivePrefix = "arXiv",
    reportNumber = "SLAC-REPRINT-1999-098, CLNS-99-1611, CLEO-99-3",
    doi = "10.1103/PhysRevD.61.052001",
    journal = "Phys. Rev. D",
    volume = "61",
    pages = "052001",
    year = "2000"
}

@article{CLEO:2007vpk,
    author = "Adam, N. E. and others",
    collaboration = "CLEO",
    title = "{A Study of Exclusive Charmless Semileptonic B Decay and $|V_{ub}|$}",
    eprint = "hep-ex/0703041",
    archivePrefix = "arXiv",
    reportNumber = "CLNS-06-1982, CLEO-06-22",
    doi = "10.1103/PhysRevLett.99.041802",
    journal = "Phys. Rev. Lett.",
    volume = "99",
    pages = "041802",
    year = "2007"
}

@article{Belle:2006hlt,
    author = "Hokuue, T. and others",
    collaboration = "Belle",
    title = "{Measurements of branching fractions and $q^2$ distributions for $B \to \pi \ell \nu$ and $B \to \rho \ell \nu$ decays with $B \to D(*) \ell \nu$ decay tagging}",
    eprint = "hep-ex/0604024",
    archivePrefix = "arXiv",
    reportNumber = "BELLE-PREPRINT-2006-10, KEK-PREPRINT-2006-4",
    doi = "10.1016/j.physletb.2007.02.067",
    journal = "Phys. Lett. B",
    volume = "648",
    pages = "139--148",
    year = "2007"
}

@article{Becirevic:2026tle,
    author = "Be{\v{c}}irevi{\'c}, D. and Martines, M. and Rosauro-Alcaraz, S. and Sumensari, O.",
    title = "{Interpreting the results on exclusive $c \ to s \mu \nu$ modes}",
    eprint = "2603.25837",
    archivePrefix = "arXiv",
    primaryClass = "hep-ph",
    doi = "10.1016/j.physletb.2026.140713",
    journal = "Phys. Lett. B",
    volume = "880",
    pages = "140713",
    year = "2026"
}

@article{Trotta_2008,
   title={\textit{Bayes in the sky: Bayesian inference and model selection in cosmology}},
   volume={49},
   ISSN={1366-5812},
   url={http://dx.doi.org/10.1080/00107510802066753},
   DOI={10.1080/00107510802066753},
   number={2},
   journal={Contemporary Physics},
   publisher={Informa UK Limited},
   author={Trotta, Roberto},
   year={2008},
   month=Mar, pages={71–104} }

@article{10.1214/ss/1015346320,
author = {Roberts, Gareth and Rosenthal, Jeffrey},
year = {2001},
month = {11},
pages = {},
title = {\textit{Optimal Scaling for Various Metropolis-Hastings Algorithms}},
volume = {16},
journal = {Statistical Science},
doi = {10.1214/ss/1015346320}
}

@misc{Wolfram:Doc,
    key= {SmoothKernelDistribution},
    title = "{Wolfram Language Documentation}",
    note = { \href{https://reference.wolfram.com/language/ref/SmoothKernelDistribution.html.en}{[reference.wolfram.com]}}
}

@article{Cornella:2026zkd,
    author = {Cornella, Claudia and Ferr{\'e}, Max and K{\"o}nig, Matthias and Neubert, Matthias},
    title = "{The Simplest B Decay, Precisely}",
    eprint = "2601.14361",
    archivePrefix = "arXiv",
    primaryClass = "hep-ph",
    reportNumber = "CERN-TH-2026-006, MITP-25-075",
    month = "1",
    year = "2026"
}

@article{Bharucha:2015bzk,
    author = "Bharucha, Aoife and Straub, David M. and Zwicky, Roman",
    title = "{$B\to V\ell^+\ell^-$ in the Standard Model from light-cone sum rules}",
    eprint = "1503.05534",
    archivePrefix = "arXiv",
    primaryClass = "hep-ph",
    reportNumber = "TUM-HEP-957-14, CP3-Origins-2015-010, DIAS-2015-10",
    doi = "10.1007/JHEP08(2016)098",
    journal = "JHEP",
    volume = "08",
    pages = "098",
    year = "2016"
}

@article{Bourrely:2008za,
    author = "Bourrely, Claude and Caprini, Irinel and Lellouch, Laurent",
    title = "{Model-independent description of B ---{\ensuremath{>}} pi l nu decays and a determination of |V(ub)|}",
    eprint = "0807.2722",
    archivePrefix = "arXiv",
    primaryClass = "hep-ph",
    reportNumber = "CPT-P36-2007",
    doi = "10.1103/PhysRevD.82.099902",
    journal = "Phys. Rev. D",
    volume = "79",
    pages = "013008",
    year = "2009",
    note = "[Erratum: Phys.Rev.D 82, 099902 (2010)]"
}

@article{Leljak:2021vte,
    author = "Leljak, Domagoj and Meli{\'c}, Bla{\v{z}}enka and van Dyk, Danny",
    title = "{The $ \overline{B} $ {\textrightarrow} {\ensuremath{\pi}} form factors from QCD and their impact on |V$_{ub}$|}",
    eprint = "2102.07233",
    archivePrefix = "arXiv",
    primaryClass = "hep-ph",
    reportNumber = "EOS-2021-02, TUM-HEP-1316/21, RBI-ThPhys-2021-1",
    doi = "10.1007/JHEP07(2021)036",
    journal = "JHEP",
    volume = "07",
    pages = "036",
    year = "2021"
}

@article{FlavourLatticeAveragingGroupFLAG:2024oxs,
    author = "Aoki, Y. and others",
    collaboration = "Flavour Lattice Averaging Group (FLAG)",
    title = "{FLAG review 2024}",
    eprint = "2411.04268",
    archivePrefix = "arXiv",
    primaryClass = "hep-lat",
    reportNumber = "CERN-TH-2024-192, FERMILAB-PUB-24-0785-T",
    doi = "10.1103/nfzp-p5dn",
    journal = "Phys. Rev. D",
    volume = "113",
    number = "1",
    pages = "014508",
    year = "2026"
}

@article{Belle-II:2024xwh,
    author = "Adachi, I. and others",
    collaboration = "Belle-II",
    title = "{Determination of |Vub| from simultaneous measurements of untagged B0{\textrightarrow}{\ensuremath{\pi}}-{\ensuremath{\ell}}+{\ensuremath{\nu}}{\ensuremath{\ell}} and B+{\textrightarrow}{\ensuremath{\rho}}0{\ensuremath{\ell}}+{\ensuremath{\nu}}{\ensuremath{\ell}} decays}",
    eprint = "2407.17403",
    archivePrefix = "arXiv",
    primaryClass = "hep-ex",
    reportNumber = "Belle II Preprint 2024-023, KEK Preprint 2024-21",
    doi = "10.1103/jnwn-ts6q",
    journal = "Phys. Rev. D",
    volume = "111",
    number = "11",
    pages = "112009",
    year = "2025"
}

@article{Grzadkowski:2010es,
    author = "Grzadkowski, B. and Iskrzynski, M. and Misiak, M. and Rosiek, J.",
    title = "{Dimension-Six Terms in the Standard Model Lagrangian}",
    eprint = "1008.4884",
    archivePrefix = "arXiv",
    primaryClass = "hep-ph",
    reportNumber = "IFT-9-2010, TTP10-35",
    doi = "10.1007/JHEP10(2010)085",
    journal = "JHEP",
    volume = "10",
    pages = "085",
    year = "2010"
}

@article{Colangelo:2019axi,
    author = "Colangelo, P. and De Fazio, F. and Loparco, F.",
    title = "{Probing New Physics with $\bar B \to \rho(770) \, \ell^- \bar \nu_\ell$ and $\bar B \to a_1(1260) \, \ell^- \bar \nu_\ell$}",
    eprint = "1906.07068",
    archivePrefix = "arXiv",
    primaryClass = "hep-ph",
    reportNumber = "BARI-TH/19-720",
    doi = "10.1103/PhysRevD.100.075037",
    journal = "Phys. Rev. D",
    volume = "100",
    number = "7",
    pages = "075037",
    year = "2019"
}

@article{Colangelo:2024mxe,
    author = "Colangelo, Pietro and De Fazio, Fulvia and Loparco, Francesco and Losacco, Nicola",
    title = "{New physics couplings from angular coefficient functions of  $B^- \to D^* (D \pi) \ell  \nu_\ell$}",
    eprint = "2401.12304",
    archivePrefix = "arXiv",
    primaryClass = "hep-ph",
    reportNumber = "BARI-TH/754-24",
    doi = "10.1103/PhysRevD.109.075047",
    journal = "Phys. Rev. D",
    volume = "109",
    pages = "075047",
    year = "2024"
}

@article{Gambino:2020jvv,
    author = "Gambino, P. and others",
    title = "{Challenges in semileptonic $B$ decays}",
    eprint = "2006.07287",
    archivePrefix = "arXiv",
    primaryClass = "hep-ph",
    reportNumber = "FERMILAB-PUB-20-235-T",
    doi = "10.1140/epjc/s10052-020-08490-x",
    journal = "Eur. Phys. J. C",
    volume = "80",
    number = "10",
    pages = "966",
    year = "2020"
}

@article{HeavyFlavorAveragingGroupHFLAV:2024ctg,
    author = "Banerjee, Sw. and others",
    collaboration = "Heavy Flavor Averaging Group (HFLAV)",
    title = "{Averages of b-hadron, c-hadron, and {\ensuremath{\tau}}-lepton properties as of 2023}",
    eprint = "2411.18639",
    archivePrefix = "arXiv",
    primaryClass = "hep-ex",
    doi = "10.1103/x87q-tld5",
    journal = "Phys. Rev. D",
    volume = "113",
    pages = "012008",
    year = "2026"
}

@article{Colangelo:2018cnj,
    author = "Colangelo, Pietro and De Fazio, Fulvia",
    title = "{Scrutinizing $ \overline{B}\to {D}^{\ast}\left(D\pi \right){\ell}^{-}{\overline{\nu}}_{\ell } $ and $ \overline{B}\to {D}^{\ast}\left(D\gamma \right){\ell}^{-}{\overline{\nu}}_{\ell } $ in search of new physics footprints}",
    eprint = "1801.10468",
    archivePrefix = "arXiv",
    primaryClass = "hep-ph",
    reportNumber = "BARI-TH-18-715",
    doi = "10.1007/JHEP06(2018)082",
    journal = "JHEP",
    volume = "06",
    pages = "082",
    year = "2018"
}

@article{Colangelo:2016ymy,
    author = "Colangelo, Pietro and De Fazio, Fulvia",
    title = "{Tension in the inclusive versus exclusive determinations of $|V_{cb}|$: a possible role of new physics}",
    eprint = "1611.07387",
    archivePrefix = "arXiv",
    primaryClass = "hep-ph",
    reportNumber = "BARI-TH-709-2016",
    doi = "10.1103/PhysRevD.95.011701",
    journal = "Phys. Rev. D",
    volume = "95",
    number = "1",
    pages = "011701",
    year = "2017"
}

@article{ParticleDataGroup:2024cfk,
    author = "Navas, S. and others",
    collaboration = "Particle Data Group",
    title = "{Review of particle physics}",
    doi = "10.1103/PhysRevD.110.030001",
    journal = "Phys. Rev. D",
    volume = "110",
    pages = "030001",
    year = "2024"
}
\end{document}